\documentclass[fleqn,usenatbib]{mnras}

\usepackage{newtxtext, newtxmath}
\usepackage{multirow}
\usepackage[T1]{fontenc}

\DeclareRobustCommand{\VAN}[3]{#2}
\let\VANthebibliography\thebibliography
\def\thebibliography{\DeclareRobustCommand{\VAN}[3]{##3}\VANthebibliography}

\usepackage{graphicx}	% Including figure files
\usepackage{amsmath}	% Advanced maths commands
\usepackage{threeparttable}
\title[LMC Geometry with multi-phase PLR]{Geometry of the LMC based on multi-phase analysis of multi-wavelength Cepheid light curves using OGLE-IV and Gaia DR3 data}

\author[G. Bhuyan et al.]{
Gautam Bhuyan$^{1}$\thanks{E-mail: gautam.bhuyan2825@gmail.com},
Sukanta Deb$^{1,2}$\thanks{E-mail: sukanta.deb@cottonuniversity.ac.in},
Shashi Kanbur$^{3}$\thanks{E-mail: shashi.kanbur@oswego.edu},
Earl P.\ Bellinger$^{4,5,6}$,
Mami Deka$^{1}$, \and
\hspace*{1mm}Anupam Bhardwaj$^{7}$ \\ 
$^{1}$Department of Physics, Cotton University, Panbazar, Guwahati 781001, Assam, India \\
$^{2}$Space and Astronomy Research Centre, Cotton University, Panbazar, Guwahati 781001, 
Assam, India \\ 
$^{3}$Department of Physics, State University of New York Oswego, Oswego, NY 13126, USA\\
$^{4}$Max Planck Institute for Astrophysics, Garching, Germany\\
$^{5}$Department of Astronomy, Yale University, CT, USA\\
$^{6}$Stellar Astrophysics Centre, Aarhus, Denmark\\
$^{7}$INAF-Osservatorio Astronomico di Capodimonte, Via Moiariello 16, 801301, Napoli, Italy }
\pubyear{2023}

\begin{document}
\label{firstpage}
\pagerange{\pageref{firstpage}--\pageref{lastpage}}
\maketitle
\begin{abstract}
The period-luminosity (PL) relation of Cepheids in the Large Magellanic Cloud (LMC) plays a pivotal role in extra-galactic distance measurement and the determination of the Hubble constant $(H_{0})$. In this work, we probe the geometry of the LMC through a detailed study of multi-phase PL relations of these Cepheids, leveraging data from the OGLE-IV and Gaia DR3 databases. We analyse the light curves of a combined sample of $\sim$3300 fundamental (FU) and first overtone (FO) mode classical Cepheids. We obtain multi-phase data with $50$ phase points over a complete pulsation cycle from the OGLE $(V, I)$ and Gaia $(G,G_{\rm BP}, G_{\rm RP})$ photometric bands. We determine the distance modulus and reddening values of individual Cepheids by fitting a simultaneous reddening law to the apparent distance modulus values. We calculate the LMC viewing angle parameters: the inclination angle $(i)$ and position angle of line of nodes $(\theta_{\rm lon})$ by fitting a plane of the form $z = f(x,y)$ to the three-dimensional distribution of Cepheids in Cartesian coordinates $(x,y,z)$. The values of LMC viewing angles from multi-phase PL relations are found to be: $i = 22\rlap{.}^{\circ}87 \pm 0\rlap{.}^{\circ}43 ~\textrm{(stat.)} \pm 0\rlap{.}^{\circ}53 ~\textrm{(syst.)}$, $\theta_{\rm lon} = 154\rlap{.}^{\circ}76 \pm 1\rlap{.}^{\circ}16 ~\textrm{(stat.)} \pm 1\rlap{.}^{\circ}01 ~\textrm{(syst.)}$, respectively. The use of multi-phase PL relations in multiple bands results in lower uncertainties for the LMC viewing angle parameters as compared to those derived from the mean light PL relations. This shows that the use of multi-phase PL relations with multi-wavelength photometry significantly improves the precision of these measurements, allowing better constraints on the morphology and the structure of the LMC. 
\end{abstract}

% Select between one and six entries from the list of approved keywords.
% Don't make up new ones.
\begin{keywords}
stars: variables: Cepheids -- Magellanic Clouds, methods: data analysis, statistical 

\end{keywords}

%%%%%%%%%%%%%%%%%%%%%%%%%%%%%%%%%%%%%%%%%%%%%%%%%%

%%%%%%%%%%%%%%%%% BODY OF PAPER %%%%%%%%%%%%%%%%%%

\section{Introduction} \label{intro}
Classical Cepheids (hereafter, Cepheids) are highly luminous Population-I stars located in the Cepheid instability strip, with luminosities up to $10^{5}~\rm{L}_{\odot}$ having typical masses $M\sim 3-10~\rm{M}_{\odot}$ \citep{subrs17,bhar20}. They belong to a class of intrinsic variable stars where the stars undergo pulsations due to perturbations from hydrostatic equilibrium being amplified by the $\kappa$ and $\gamma$ mechanisms in the partial ionization zones \citep{cate15,bhar20}. Cepheids are used as ``standard candles'' because they obey a well-defined period-luminosity (PL) relation, also known as the Leavitt law, in which the luminosity of the star increases with increasing period \citep{leavitt12}. Calibration of Cepheid PL relations in the Magellanic Cloud (MC) dwarf galaxies is essential to establish the extra-galactic distance scales and determine the present value of the Hubble constant, which tells us the current expansion rate of the universe \citep{free01, free12,ries16,ries22}. In addition to calibrating the Leavitt law \citep{ngeo12,ngeo22,ripe23}, its sensitivity to Cepheid metallicity \citep{roma08,roma22,gier18,breu22}, the non-linearity in PL relations \citep{ngeo06, ngeo08, kanb06, bhar16} as well as the universality of the PL relations \citep{bono10} have been extensively studied.% in the literature. 

Majority of the studies in the literature have derived PL relations based on mean magnitudes of Cepheids. The PL relations of Cepheids are affected by the nature of their period-colour (PC) relations through the period-luminosity-colour (PLC) relation \citep{kanb96,kanb04}. Cepheids in the LMC have been found to follow distinct PC relations at the phases of maximum, minimum, and mean light by several studies in the literature \citep{kanb04ii,kanb06,kanb07,bhar14,das20}. Extensive studies on PC and PL relations of Cepheids in the LMC have suggested the existence of a significant break at $P = 10$ days for FU-mode \citep{ngeo08} and at $P = 2.5$ days for FO-mode Cepheids \citep{bhar14, bhar16}. These findings have led to various theoretical and empirical investigations on multi-phase PL relations \citep{ngeo06, kanb09, ngeo12, kurb23}.

The Large Magellanic Cloud (LMC) also serves as a laboratory in understanding the geometry, structure, and evolution of galaxies as well as for resolved stellar population studies. The LMC is considered to have roughly a planar geometry \citep{mare01}, with an inclined and rotating star forming disk \citep{mare02, subrs10}. The LMC has a spiral arm and a central bar region which is off-centred \citep{subrs09}, likely due to tidal interactions with the Small Magellanic Cloud (SMC) and the Milky Way (MW) galaxy \citep{west97, choi18, jime23}. The galaxy hosts several star clusters \citep{song21}; active star-forming regions such as the Tarantula Nebula \citep[$30$ Doradus,][]{evan11, tatt13, fahr23}; and the remnants of type-\textsc{II} supernova \textsc{SN1987A} \citep{west87A}. 

The geometry of a galaxy is characterized by its angular orientation and axis ratios in the three different directions. The orientation of a galaxy is obtained in the form of its viewing angles, viz., inclination angle $(i)$ and position angle of line of nodes $(\theta_{\rm lon})$. The geometrical and structural features of the LMC have been studied using different kinds of tracers such as the core He-burning Red Clump (RC) stars based on their colour-magnitude diagram (CMD) \citep{olse02, subra03, subrs10, choi18} and variable stars like Cepheids relying on mean light PL or period-Wesenheit (PW) relations (\citealt{niko04, inno16, deb18}, to name a few). However, no such study on the geometry of the LMC has been carried out using multi-phase PL relations of Cepheid variables. 

% This serves as a motivation to study the geometry of the LMC using multi-phase PL relations.     
This study deals with the determination of the LMC viewing angle parameters for the first time by simultaneously using  the multi-phase PL relations and the multi-wavelength photometry based on the LMC Cepheid light curves. The availability of data generated from OGLE-IV experiment and Gaia space telescope data release 3 (Gaia DR3) with complete phase coverage enables us to study the Cepheid multi-phase PL relations using multi-wavelength photometry. This also provides a unique opportunity to determine the viewing angle parameters of the LMC. The data and methodology used in this paper are described in Section \ref{data}. The results of multi-phase PL relations as well as the determination of LMC viewing angles are presented in Section \ref{result}. Finally, the summary and conclusion of the study are presented in the Section \ref{summ}. 

\section{Data and Methodology} 
\label{data}
\subsection{Data} 
The fourth phase of the Optical Gravitational Lensing Experiment (OGLE-IV) project \citep{sosz15} provides an up-to-date ground-based optical photometry of Cepheids, other type of variable stars as well as several types of celestial objects in the Milky Way, LMC and SMC. The OGLE-IV archival database contains light curve data of $2477$ Fundamental (FU) mode and $1776$ first-overtone (FO) mode Cepheids in the LMC  with more than $99\%$ photometric completeness. The OGLE-IV Cepheids are then cross-matched with Gaia DR3 \citep{gaia22b} database using the CDS-Xmatch service\footnote{\url{http://cdsxmatch.u-strasbg.fr/}} to retrieve the photometric data in Gaia photometric bands. Gaia DR3 is the latest large-scale database available from the Gaia spacecraft. Cross-matching with the Gaia data results in $2252$ FU-mode and $1480$ FO-mode Cepheids in Gaia bands.
\subsection{Fourier Decomposition of Cepheid Light curves}  \label{method1}
 The individual light curves of Cepheids are folded into phase using  \citep{deb09}:
\begin{align}
\Phi &= \frac{t - t_{0}}{P} - \rm{Int} {\left(\frac{t - t_{0}}{P} \right)}.
\label{eq:1}
\end{align}
Here $P$ denotes period of the star in days, $t$ denotes time of observation $t_{0}$ denotes the epoch of maximum light, and \textit{Int} denotes the integer part of the quantity. The values of $\Phi$ ranges from 0 to 1 corresponding to a complete pulsation cycle. The phased light curves of Cepheids are then fitted with a cosine Fourier function of the form:
\begin{align}
m(t) &= A_{0} + \sum_{i=1}^{N} A_{i} \cos{(i\omega(t-t_{0}) + \phi_{i})}. 
\label{eq:2}
\end{align}
Here $A_{i}$ and $\phi_{i}$ denote the Fourier coefficients, where $i=1,2,\dots,N$. To fit the observed Cepheid light curves with the Fourier expansion in Equation~\eqref{eq:2}, there  are $2N+1$ unknown parameters to be determined. This requires the observed light curves to have at least $2N+1$ data points. The optimal value of $N$, is obtained using the Baart's criterion \citep{peter86}. Most of the OGLE-IV Cepheid light curves have well-sampled data points with complete phase coverage. To avoid any under-sampling and ensure sufficient number of data points in a complete pulsation cycle, Cepheids with at least $30$ observations in both $V$ \& $I$ bands are selected for the Fourier fitting of light curves. The Gaia band observations are sparse as compared to the OGLE data. Hence, Cepheid light curves with at least $10$ observations in the Gaia bands are chosen for the Fourier fitting. Multi-phase values of magnitudes of the Cepheid light curves at $50$ different phase points are then obtained using the Fourier coefficients of the fitted light curves.

\begin{figure}
    \centering
    \includegraphics[width=0.95\linewidth, keepaspectratio, trim={0.3in 0.6in 0 0}, clip]{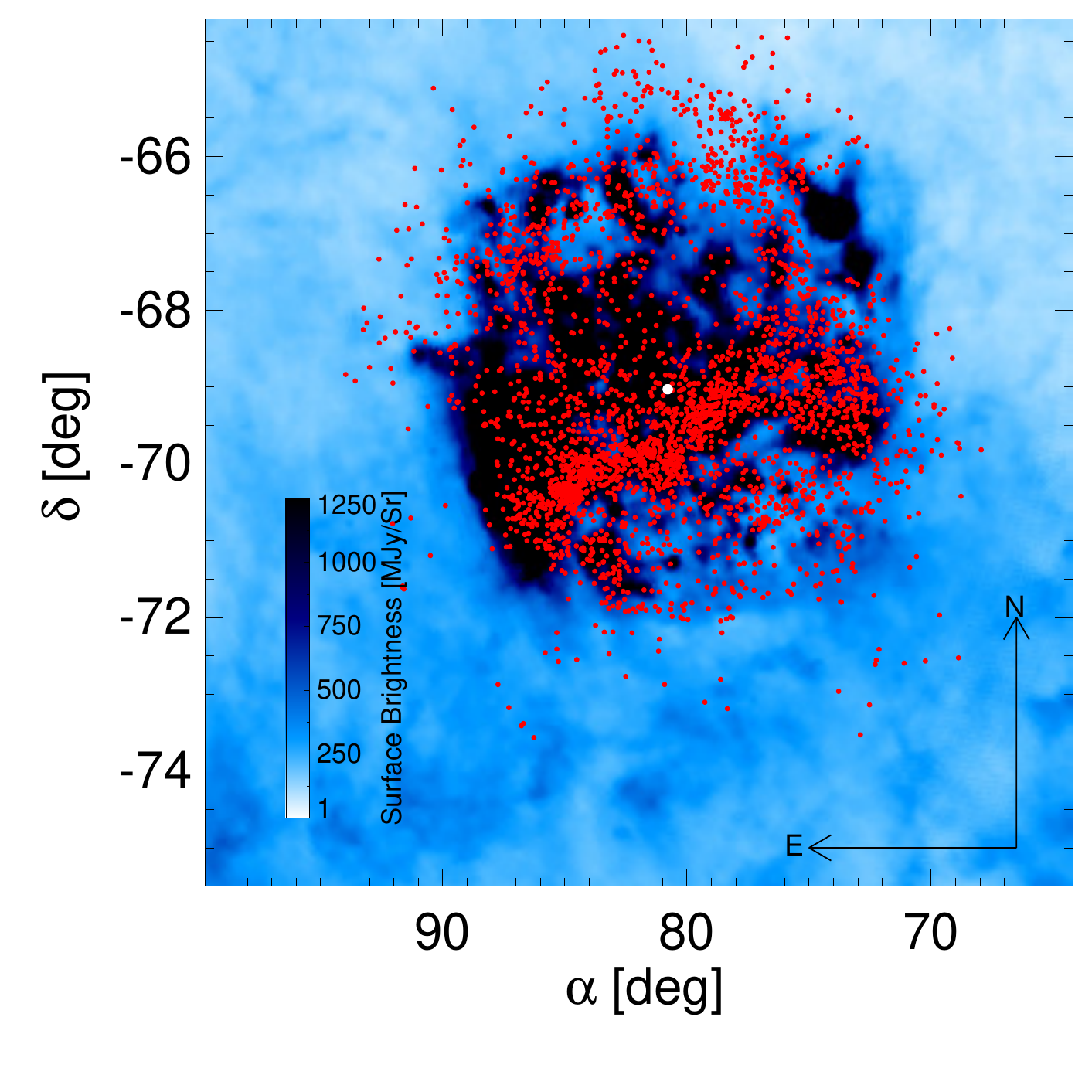}
    \caption{Distribution of the sample of $3293$ classical Cepheids shown as red dots is over-plotted on the false-colour IRAS image of the LMC transformed from a grey-scale image. The surface brightness expressed in $\rm MJySr^{-1}$  is shown in a color-bar plot. The centre of the LMC, $(\alpha_{0}, \delta_{0})$ = $(80.78, -69.03)$ \citep{niko04} adopted  in this study is marked with a bigger filled white circle.}
    \label{fig:fig1}
\end{figure}

The final light curve data contain $2007$ FU-mode and $1286$ FO-mode common Cepheids (hereafter, common sample) in all the five bands, making a combined (FU+FO) sample of $3293$ Cepheids. Fig.\ref{fig:fig1} shows the distribution of the common sample of Cepheids in the LMC in equatorial coordinates (RA-DEC). This distribution is plotted on the Infrared Astronomical Satellite (IRAS) FITS image of size $(300 \times 300)$ pixels covering $\sim (11.87 \times 11.87)$ square degrees of the sky. It is obtained from Schlegel, Finkbeiner and Davis \textsc{(SFD)} 100-micron survey using SkyView\footnote{\url{https://skyview.gsfc.nasa.gov/current/cgi/query.pl}} interface \citep{mcgl98}. A false-colour image denoting the surface brightness in $\rm MJy Sr^{-1}$ units at different parts of the LMC is obtained from the grey-scale FITS image.

\subsection{Multi-phase PL Relations, Distances and Reddening} \label{method2}
 The use of multi-phase PL relations to determine the distances and the reddening values of the LMC Cepheids is motivated by the existence of the PL relation with minimum scatter at a particular phase \citep{bhar19,kurb23}. Furthermore, the advantages of using multi-wavelength photometry in the extra-galactic distances have been briefly outlined in \citet{deb18}. Simultaneous determination of distance and reddening values of the LMC Cepheids using multi-phase PL relations based on multi-wavelength photometry performed in this study will lead to their independent values free from any systematic effect arising due to the dependency of these relations on metallicity. This in turn will help in the accurate and precise determination of the geometry of the galaxy. Multi-phase PL relations of classical Cepheids over a complete pulsation cycle are studied using five photometric bands making use of OGLE-IV and Gaia DR3 data. Mathematically, the pulsation periods of Cepheids are related to their intrinsic luminosities or absolute magnitudes through the relation \citep{niko04}:
\begin{align}
M_{\lambda} &= \alpha_{\lambda} + \beta_{\lambda} \log{P} + \epsilon_{\lambda}(M, T_{e}, Z...). \label{eq:3} 
\end{align}
Here  $M_{\lambda}$ represents the absolute magnitude of a Cepheid in a particular photometric band $\lambda$. The parameters $\alpha_{\lambda}$ and $\beta_{\lambda}$ represent the slopes and intercepts of the PL relations, respectively, for the photometric band $\lambda$. The term $\epsilon_{\lambda}(M, T_{e}, Z...)$ takes care of any unknown contribution to the PL relation due to the variation in Cepheid metallicity $(Z)$, effective temperature $(T_{\rm eff})$ or masses $(M)$. The apparent distance modulus in a particular photometric band $\mu_{\lambda}$ is given by:
\begin{align}
\mu_{\lambda} = & \mu_{0} + A_{\lambda}, \label{eq:4}
 \end{align}
where $\mu_{0}$ denotes the true distance modulus and  $A_{\lambda} = R_{\lambda} E(B-V)$, is the interstellar extinction in the photometric band $\lambda$. $R_{\lambda}$ is called the ratio of total to selective absorption in the band $\lambda$; $E(B-V)$ represents individual reddening values of Cepheids. The central wavelengths of the five photometric bands $V, I, G, G_{\rm BP}$, and $G_{\rm RP}$ used in the analysis, are approximately taken as: $\lambda_{V}=0.556~\micron$, $\lambda_{I}=0.810~\micron$, $\lambda_{G}=0.673~\micron$, $\lambda_{G_{\rm BP}} = 0.532~\micron$ and $\lambda_{G_{\rm RP}}=0.797~\micron$, respectively \citep{bona10,jord10, wang19,iwan22}. The $R_{\lambda}$ values corresponding to these wavelengths are found to be: $R_{\lambda}=(3.23, 1.97, 2.62, 3.40, 2.03)$, by applying the \citet{ccm89} reddening law using a fixed value of $R_{V}=3.23$. For comparison with the \citet{skow21} reddening map, the reddening values $E(B-V)$ are converted into their corresponding $E(V-I)$ values. The scaling relation $E(V-I) = 1.26 ~E(B-V)$, suitable for the $R_{V}=3.23$ reddening law, is used \citep{kurb23}. Putting $\mu_{\lambda}$ into the Equation~\eqref{eq:4}, it can be written and rearranged into the following relation:

\begin{align}
m_{\lambda} = & \alpha_{\lambda} + \beta_{\lambda} \log{P}+ \mu_{0} + R_{\lambda}E(B-V) + \epsilon_{\lambda}(M, T_{e}, Z...). 
\label{eq:5}
\end{align}
The true distance modulus $\mu_{0}$ and absolute reddening $E(B-V)$ of the $i^{\rm th}$ Cepheid are determined using  \citep{niko04, deb18}:
\begin{align}
\mu_{0,i} = & \overline{\mu}_{\tiny LMC} + \Delta \mu_{0,i}, \nonumber \\ 
E(B-V)_{i} = & \overline{E(B-V)}_{\tiny LMC} + \Delta E(B-V)_{i}. 
\label{eq:6}
\end{align}
Here $\overline{\mu}_{\tiny LMC}$ and $\overline{E(B-V)}_{\tiny LMC}$ denote the mean distance and average reddening value of the LMC, taken from the literature. Therefore, the apparent magnitude, $m_{\lambda,i,j}$ of the $i^{\rm th}$ Cepheid in its $j^{\rm th}$ pulsation phase for the band $\lambda$ can be written as:
\begin{align}
m_{\lambda,i,j} = & \overline{\mu}_{LMC} + \overline{E(B-V)}_{LMC} + \nonumber
\epsilon_{\lambda}(M, T_{e}, Z...) \nonumber +  \\
&\alpha_{\lambda,j} + \beta_{\lambda,j} \log{P_{i}}+ \Delta\mu_{0,i} + \nonumber 
R_{\lambda,j}\Delta E(B-V)_{i} \nonumber \\
\rm{Or}~~ m_{\lambda,i,j} = & \alpha^{\prime}_{\lambda,j} + \beta_{\lambda,j} \log{P_{i}}+ \Delta\mu_{0,i} + R_{\lambda,j}\Delta E(B-V)_{i}. \label{eq:7}
\end{align}
The coefficient $\alpha_{\lambda,j}^{\prime}$ represents the modified zero-point of the PL relation into which $\overline{\mu}_{LMC}$, $\overline{E(B-V)}_{LMC}$ and $\epsilon_{\lambda}(M, T_{e}, Z...)$ are subsumed. Equation~\eqref{eq:7} can be solved following \cite{niko04} and \cite{deb18} in two iterative steps. However, the mathematical formulations used in this study are different from  the previous two and involve simultaneous fitting applying linear algebraic methods. In the first iteration, the coefficients $\alpha_{\lambda,j}^{\prime}$ and $\beta_{\lambda,j}$ are obtained using:
\begin{align}
m_{\lambda,i,j} \approx \alpha_{\lambda,j}^{\prime} + \beta_{\lambda,j}\log{P_{i}}.
\label{eq:8}
\end{align}
Here $m_{\lambda,i,j}$ represents apparent magnitude at the $j^{\rm th}$ phase of an $i^{\rm th}$ Cepheid. The coefficients are obtained by carrying out a simultaneous fit to the multi-band and multi-phase Cepheid light curve data using a generalized linear model (GLM)\footnote{\url{https://timeseriesreasoning.com/contents/deep-dive-into-variance-covariance-matrices/}} of the following form \citep{green00}:
\begin{align}
{\mathbf y} &= \mathbf{X q} + {\mathbf \xi}. \label{eq:9}    
\end{align}
The multi-phase magnitudes of Cepheids in the OGLE and Gaia bands constitute the dependent variable $\mathbf{y}$. The matrix of the independent variables, $\mathbf{X}$ is the design matrix constructed to simultaneously fit the multi-phase light curves to obtain the multi-phase PL relations in the form of the parameters $\mathbf{\hat{q}}$. The form of $\mathbf{X}$ is shown in the Appendix~\ref{appendix}. Here the $\mathbf{q}$ vector contains the multi-phase PL parameters $(\alpha_{\lambda,j}^{\prime}$, $\beta_{\lambda,j})$ to be estimated and $\xi$ is a vector containing parameter errors. The best fit values of $\mathbf{q}$ obtained from GLM are denoted by $\mathbf{\hat{q}}$. In the first iteration, ordinary least squares (OLS) minimization technique is used to obtain the best fit parameters $\mathbf{\hat{q}}$ by minimizing the residual sum of squares (RSS) $(\bf{y} - \bf{X} {\bf \hat{q}})^{T} (\bf{y} - \bf{X} {\bf \hat{q}})$. The parameter values ${\bf \hat{q}}$ are obtained using:
\begin{align}
{\bf \hat{q}} &= ({\bf X^{T} X})^{-1} {\bf X^{T} y}. \label{eq:10} 
\end{align}
The corresponding statistical errors associated with these parameters are calculated using variance-covariance matrix of the GLM:
\begin{align}
\textrm{Var}(\hat{q}) &= \frac{\bf e^{T}e}{(n-k)} ({\bf X^{T} X})^{-1}, \label{eq:11}
\end{align}
where ${\bf e} = (\bf{y} - \bf{X} {\bf \hat{q}})$; $n$ denotes the sample size; $k$ denotes the number of parameters to be estimated. Once the values of  $(\alpha_{\lambda,j}^{\prime}$, $\beta_{\lambda,j})$ are obtained from the first iteration, the true individual distance modulus and reddening values of each star with respect to the mean values of the LMC are obtained as follows:
\begin{align}
\Delta\mu_{\lambda,i,j} &= m_{\lambda,i,j} - (\alpha_{\lambda,j}^{\prime} + \beta_{\lambda,j} 
\log{P_{i}}) \nonumber \\ 
\Delta \mu_{\lambda,i,j} &= \Delta \mu_{0,i} + R_{\lambda,j} \Delta E(B-V)_{i}. \label{eq:12}
\end{align}
The apparent distance moduli, $\Delta\mu_{\lambda,i,j}$ of Cepheids in multi-phase constitute the dependent variable $\mathbf{y}$ in this step. The independent variable $\mathbf{X}$ contains the linear model, as given in Equation~\eqref{eq:12}. The weighted least squares (WLS) minimization technique is applied in the second iteration to obtain the values of $\Delta \mu_{0,i}$ and $\Delta E(B-V)_{i}$. The residuals $(\bf{y} - \bf{X} {\bf \hat{q}})$ are weighted in this step as the reciprocal of $\sigma_{\lambda,i,j}^{2}$, where:
\begin{align}
\sigma_{\lambda,i,j}^{2} = \sigma_{0}^{2} + \sigma_{\lambda,i,j}^{2}(\rm phot) + \sigma_{\alpha^{\prime}_{\lambda,j}}^{2} + \sigma_{\beta_{\lambda,j}}^{2} (\log{P_{i}})^{2}. \label{eq:13}
\end{align}
Here $\sigma_{0}$ and $\sigma_{\lambda,i,j}(\rm phot)$ denote the intrinsic and photometric errors, respectively. The quantities $\sigma_{\alpha^{\prime}_{\lambda,j}}$ and $\sigma_{\beta_{\lambda,j}}$ represents errors in the parameters  $\alpha^{\prime}_{\lambda,j}$ and $\beta_{\lambda,j}$. The photometric errors are determined by using Monte Carlo simulations \citep{deb15} with 100 iterations. In each iteration, Gaussian noise is added to the observed light curves and fitted magnitudes are obtained using Fourier decomposition. This results in bootstrap simulations of the fitted magnitudes. We construct the variance-covariance matrix of the fitted magnitudes of each light curve using:
\begin{align}
{\rm Cov}(Y_{i}, Y_{j}) = \frac{\sum_{l, k = 1}^{N_{\rm iter}} (Y_{i,l} - \overline{Y_{i}})(Y_{j,k} - \overline{Y_{j}})}{(N_{\rm iter}-1)}. \label{eq:14}
\end{align}
Here $\overline{Y_{i}}$, $\overline{Y_{j}}$ represent the mean values of bootstrap simulations of the fitted magnitudes at the phase points $i$ and $j$, respectively. $N_{\rm iter}$ denotes the number of bootstrap simulations. From the diagonal elements of ${\rm Cov}(Y_{i}, Y_{j})$, corresponding errors in the fitted magnitudes are calculated. We adopt $\sigma_{0} = 0.05$ following \citet{niko04} and \citet{deb18} for the analysis. The best fit parameters ${\bf \hat{q}}$ are obtained by minimizing the weighted residual sum of squares: $(\bf{y} - \bf{X} {\bf \hat{q}})^{T} C (\bf{y} - \bf{X} {\bf\hat{q}})$. The parameter matrix can be obtained as follows:
\begin{align}
{\bf \hat{q}} &= ({\bf X^{T} C X})^{-1} {\bf X^{T} C y}, \label{eq:15}
\end{align}
where ${\bf C}$ represents a diagonal matrix with $1/\sigma_{\lambda,i,j}^{2}$ as the diagonal elements. The statistical errors in the values of $\Delta \mu_{0,i}$ and $\Delta E(B-V)_{i}$ are estimated using:
\begin{align}
\rm{Var}(\hat{q}) &= \frac{\bf e^{T} C ~e}{(n-k)} ({\bf X^{T} C X})^{-1}. \label{eq:16}
\end{align}
The multi-phase magnitudes of the LMC Cepheids are corrected for distance and reddening using the determined values of $\Delta \mu_{0,i}$ and $\Delta E(B-V)_{i}$. The multi-phase PL relations thus obtained in the second iteration show reduced dispersion. 

We also investigate the possible non-linearity in the PL relations of FU \& FO-mode Cepheids using GLM. The PL breaks are considered at $P=10$ d and $2.5$ d for the FU \citep{kanb04} and FO-mode Cepheids \citep{bhar16}, respectively. The FU-mode Cepheids with $P \geq 10$ d are referred to as the long period FU Cepheids, while those with $P<10$ d as short period FU Cepheids. Similarly, the division between the long and short period FO Cepheids is done based on $P=2.5$ d. The combined model for both long and short period FU/FO-mode Cepheids can be expressed as:
\begin{align}
m_{\lambda,i,j} \approx a (\alpha_{\lambda,j,l}^{\prime} + \beta_{\lambda,j,l}\log{P_{i}}) +
b (\alpha_{\lambda,j,s}^{\prime} + \beta_{\lambda,j,s}\log{P_{i}}). \label{eq:17} 
\end{align}
The two sets ($a=1, b=0$) and ($a=0, b=1$) correspond to long- and short-period Cepheids, respectively. The PL coefficients for long- and short-period Cepheids are denoted by $(\alpha_{\lambda,j,l}^{\prime}$, $ \beta_{\lambda,j,l})$ and $(\alpha_{\lambda,j,s}^{\prime}$, $\beta_{\lambda,j,s})$, respectively.

\section{Results} \label{result}
\subsection{Multi-phase Period-Luminosity Relations}
\begin{figure*}
     \centering
     \includegraphics[width=1\textwidth, keepaspectratio]{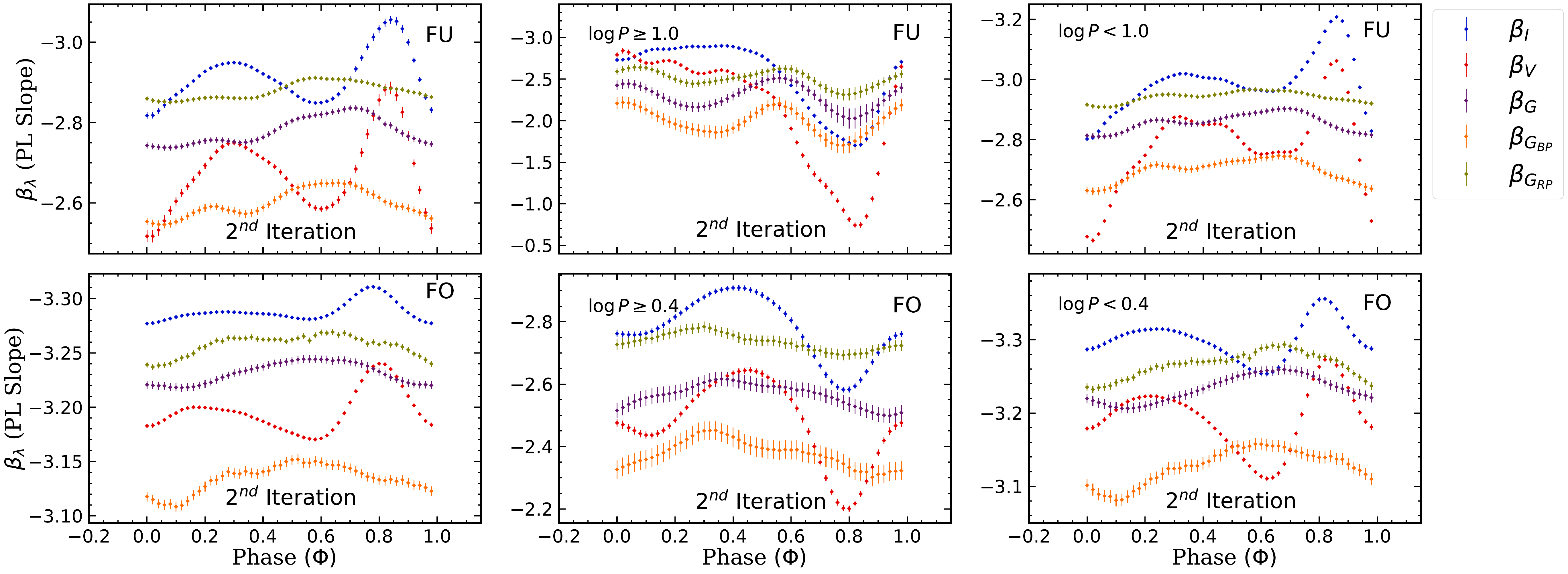}
     \caption{Variation of multi-phase PL slopes $(\beta_{\lambda})$ in OGLE-IV $(V,I)$ bands and Gaia bands ($G$, $G_{\rm BP}$,$G_{\rm RP}$) as a function of  pulsation phase $(\Phi)$ for FU and FO-mode Cepheids. Upper and lower panels show the PL slopes obtained in the second iteration for FU and FO Cepheids, respectively. Results obtained from linear PL relations are shown in the \textit{left} panels. \textit{Middle} and \textit{right} panels show  the non-linear PL relations obtained for both long and short period FU/FO-mode Cepheids, respectively. The error-bars represent statistical errors in the PL slopes, which get significantly reduced after the second iteration. The variation of multi-phase PL slopes for the long and short-period FU/FO-mode Cepheids show contrasting trends. The multi-phase linear PL slopes of FU/FO-mode Cepheids follow the same trend as that of the short period Cepheids due to their larger number as compared to the long period counterparts.}
     \label{fig:fig2}
\end{figure*}

The multi-phase study of Cepheids in the LMC using data from OGLE and Gaia bands offers some interesting results. The combined Cepheid data are fitted simultaneously with models that allow for both the existence of non-linear and purely linear relations for FU and FO-mode Cepheids. The results of multi-phase PL slopes for the FU and FO-mode Cepheids in the common sample using $50$ phase points are shown in the Fig.\ref{fig:fig2} and Fig.\ref{fig:a1}, respectively. The slopes of multi-phase PL relations obtained considering linear and non-linear PL relations are presented in separate panels in these figures. The uncertainties in the coefficients of multi-phase PL relations are found to be significantly reduced after the second iteration. The slopes of multi-phase PL relations are found to vary dynamically over the phase range $0 \leq \Phi \leq 1.0$ in all the five photometric bands. 

The observed variation of PL slopes with pulsation phase $(\Phi)$ may be attributed to the way the hydrogen ionization front (HIF) of pulsating variable stars interacts  with the photosphere during their motion through the bulk of the star \citep{skm93,kanb06, ngeo17, das20, deka22}. The middle and the left-most panels in the Fig.\ref{fig:fig2} also show that the PL slopes display contrasting trends with phase $(\Phi)$ for long as well as short-period Cepheids in both modes. This supports the findings of \citet{kurb23}.

The dispersion in multi-phase PL relations for all the five bands are presented in Fig.\ref{fig:a2}. It is quite evident from the figure that the dispersion in the PL relations are reduced significantly after second iteration in all photometric bands using the reddening values obtained in this study. However, PL dispersion in OGLE $V$- and $I$-band are reduced to a greater extent as compared to those in the Gaia bands. The results of  multi-phase Cepheid PL relations in the OGLE-IV bands are consistent with those in the literature \citep{kanb09,ngeo12,kurb23}. 

The existence of breaks in the Cepheid PL relation is examined using statistical $F$-test \citep{kanb04,bhar14} performed on the multi-phase PL relations. The results of $F$-test for 50 phase points are shown in the Fig.\ref{fig:a3}, which clearly confirms the presence of PL breaks at the specified periods. 

\subsection{LMC Reddening Map}
\begin{figure}
    \centering
    \includegraphics[width=0.45\textwidth, keepaspectratio]{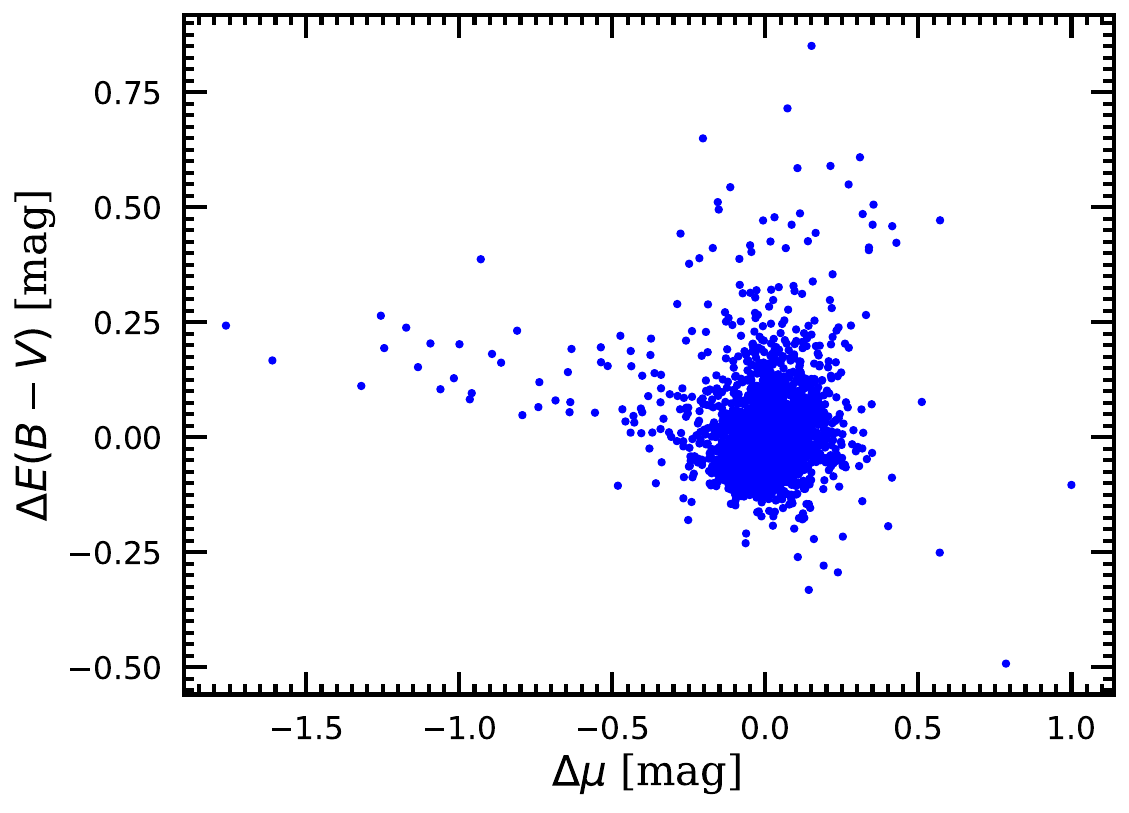}
    \caption{$\Delta E(B-V)$ versus $\Delta \mu_{0}$ plot obtained from  non-linear multi-phase PL relations with $50$ phase points for a complete pulsation cycle. Randomness in the distribution indicates that these values are uncorrelated.}
    \label{fig:fig3}
\end{figure}

The values of $\Delta\mu_{0}$ and $\Delta E(V-I)$ for each individual Cepheids in the LMC are obtained in the second iteration by carrying out a simultaneous reddening law fit to the multi-band apparent distance modulus values at all phases. The values of $\Delta \mu_{0}$ and $\Delta E(B-V)$ obtained from the simultaneous fit as shown in Fig.~\ref{fig:fig3} demonstrates them to be randomly scattered in the plot and are uncorrelated. Absolute values of distance modulus and reddening are obtained by considering the mean distance modulus and average reddening for LMC to be: $\overline{\mu}_{\rm LMC}=18.477 \pm 0.004 ({\rm stat.}) \pm 0.026 ({\rm syst.})$ mag \citep{pie19} and $\overline{E(B-V)}_{\rm LMC}= 0.14 \pm 0.02$ mag \citep{niko04}, respectively. The absolute reddening values obtained in this way yield positive values for most of the Cepheids in the present study. There are $36$ Cepheids found to have negative reddening values, which  constitutes $\sim$1.0$\%$ of all the Cepheids used in this study. However, this result is not unique to this study. Many earlier studies have also reported unphysical negative reddening values based on different tracers \citep{hasc11,deb18,mura18,pie19}. Negative reddening values may arise due to various uncertainties/propagated uncertainties of the parameters involved in the calculations \citep{mura18}. To avoid skewing the distribution towards positive reddening values and bias in the analysis, we take into account the negative reddening values.

\begin{figure*}
     \centering
     \begin{tabular}{cc}
          \resizebox{0.45\textwidth}{!}{
          \includegraphics[width=0.4\textwidth, keepaspectratio, trim={0.2in 0.6in 0 0}, clip]{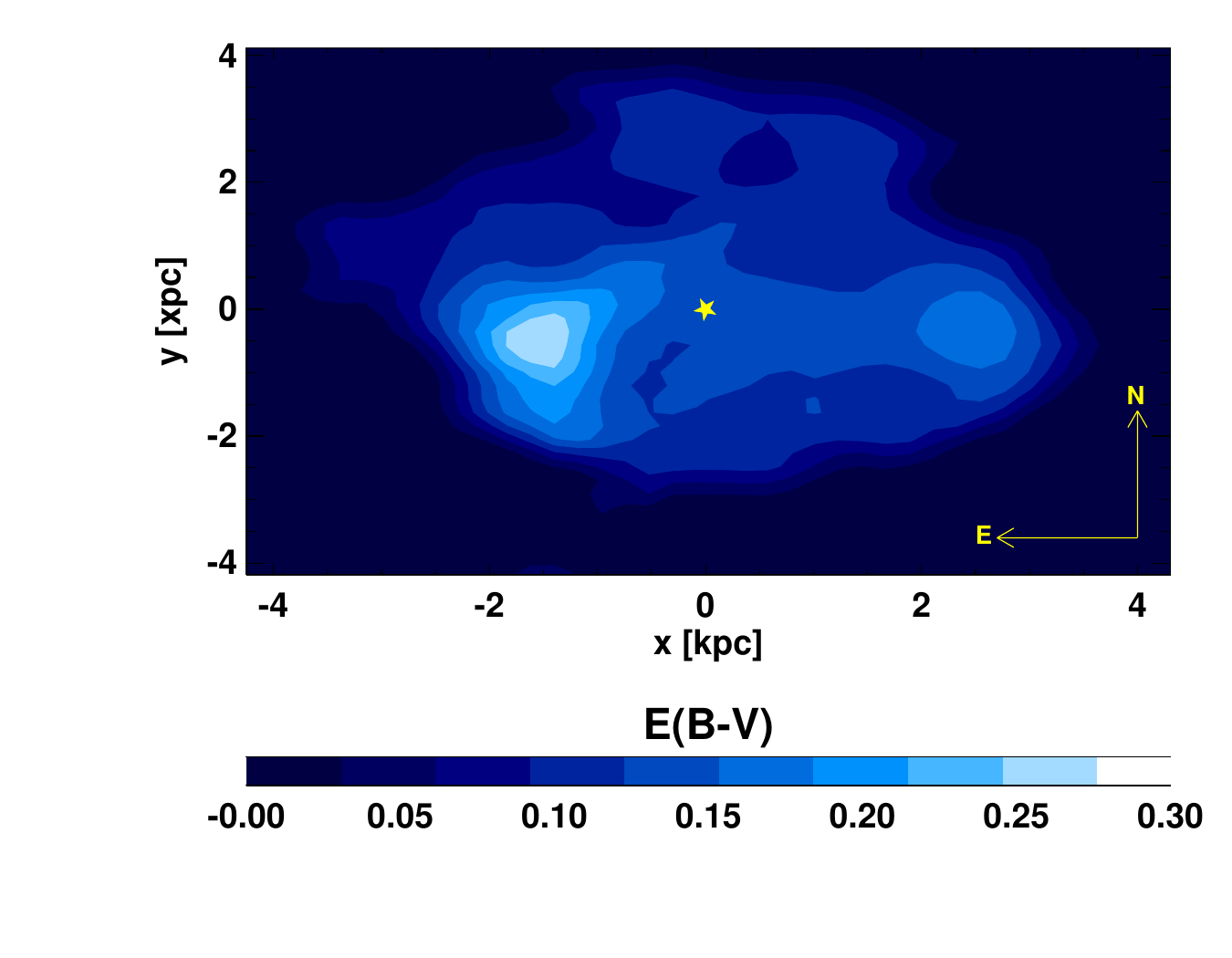}}
          % \caption{}}
          % \label{fig:fig6}}
          \resizebox{0.45\textwidth}{!}{
          \includegraphics[width=0.4\textwidth, keepaspectratio, trim={0.2in 0.6in 0 0}, clip]{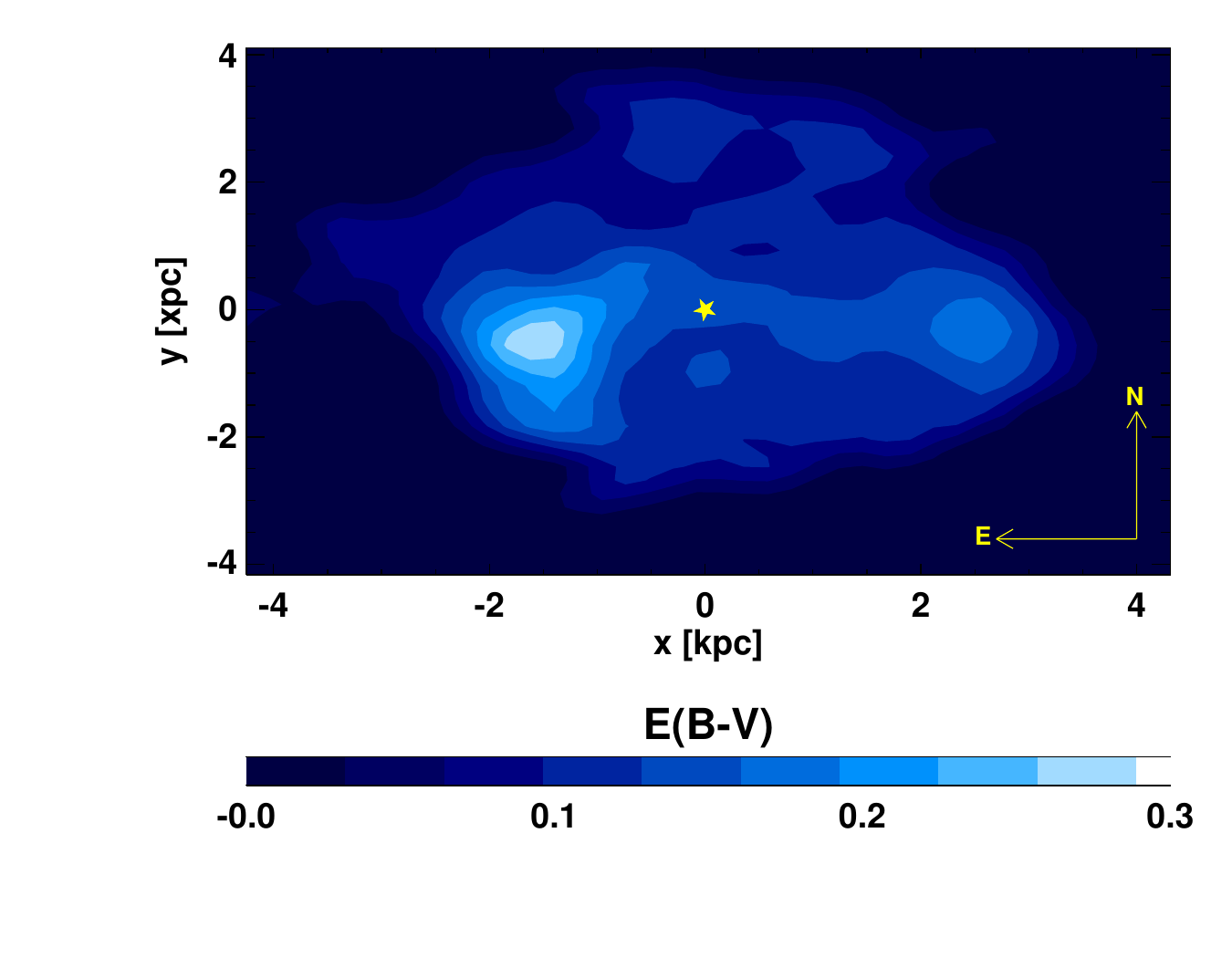}}
    \end{tabular}
    \caption{Reddening map of the LMC obtained from the multi-phase PL relations with $50$ phase points using the combined sample of FU and FO-mode Cepheids over a complete pulsation cycle. \textit{Left} panel shows the reddening map obtained when no PL break is considered. \textit{Right} panel depicts the reddening map obtained when a PL break is considered. The centre of the LMC is shown with a star symbol in yellow colour.}
    \label{fig:fig4} 
\end{figure*}

The reddening map of the LMC is constructed by using the absolute reddening values $E(B-V)$  with the RA-DEC $(\alpha, \delta)$ values as available in the OGLE-IV database\footnote{\url{http://ftp.astrouw.edu.pl/ogle/ogle4/OCVS/lmc/cep/}}. Two separate reddening maps are constructed using the results obtained from the PL relations: one considering a break and the other without a break. The resulting reddening maps for the combined sample of (FU+FO) mode Cepheids using 50 phase points are presented in Fig.\ref{fig:fig4}. Both the reddening maps of the LMC constructed using multi-phase PL relations considering break/without-break exhibit similar kinds of features. The reddening values estimated using the multi-phase PL relations with/without breaks approach a maximum of $E(B-V)=0.30$ in the eastern-most part of both the reddening maps. This region is also characterized by highest reddening values and is likely to be associated with the $30$ Doradus region (Tarantula Nebula). Tarantula Nebula is one of the closest and highly active star forming \textsc{H-II} regions in the local group \citep{evan11, tatt13, fahr23}. The same region was also found in other studies using mean light PL relations of Cepheids \citep{niko04, inno16, deb18, josh19}. 

\begin{figure*}
     \centering
     \begin{tabular}{cc}
          \resizebox{0.45\textwidth}{!}{
          \includegraphics[width=0.4\textwidth, keepaspectratio]{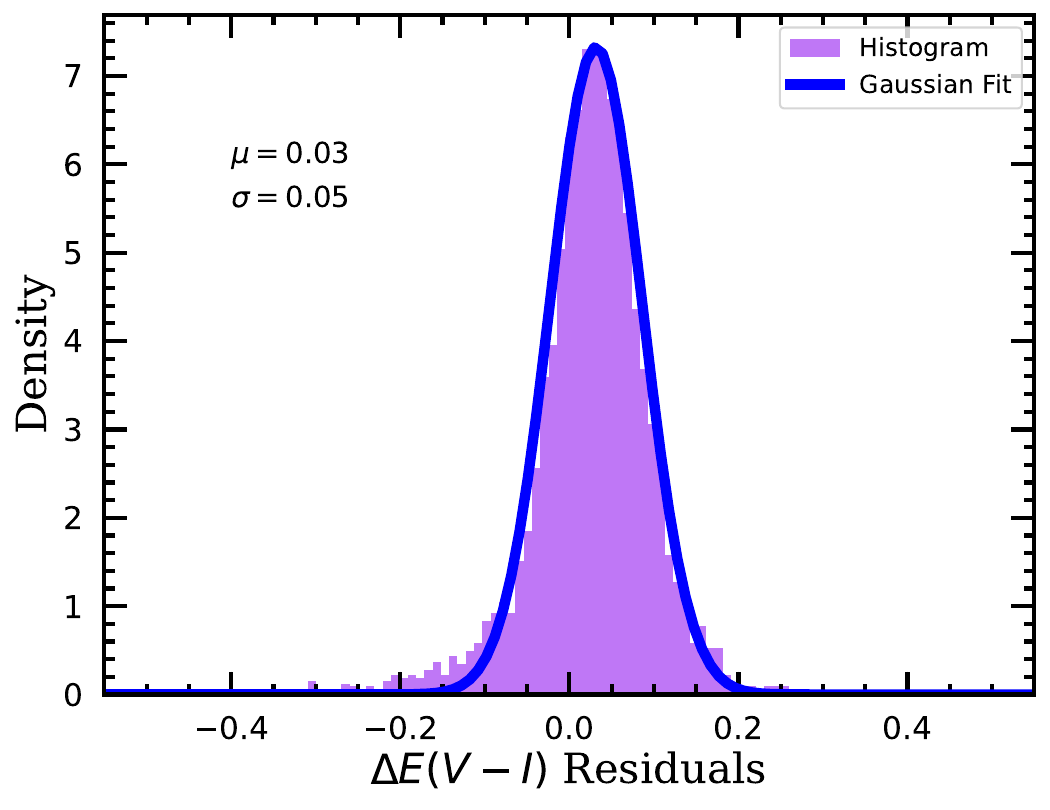}}
          % \caption{}}
          % \label{fig:fig6}}
          \resizebox{0.45\textwidth}{!}{
          \includegraphics[width=0.4\textwidth, keepaspectratio]{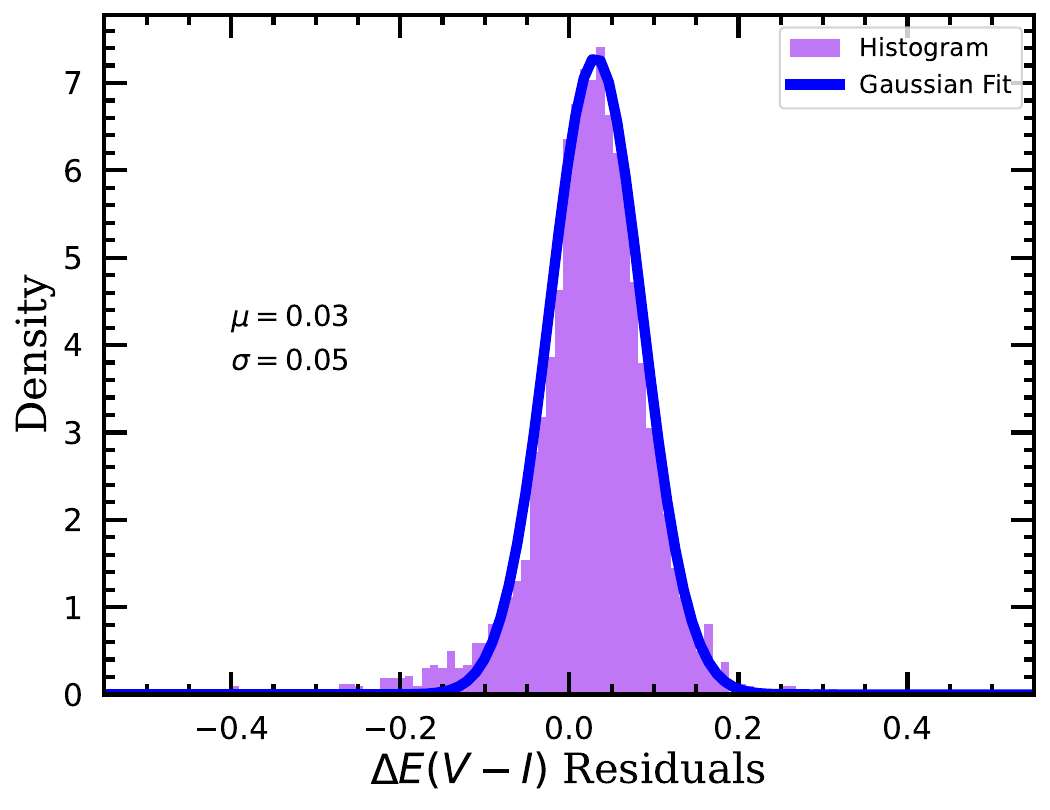}}
    \end{tabular}
    \caption{Histogram plot of the residuals between the reddening values obtained in this study and those obtained from \citet{skow21} (see text) with the mean subtracted. \textit{Left} and \textit{right} panels show the corresponding plots considering linear and non-linear PL relations, respectively. The histograms are fitted with a Gaussian to find the mean and dispersion of the residuals. 
    A small offset of $0.03$ between the two reddening values is clearly discernible.} \label{fig:fig5} 
\end{figure*}

Earlier studies have made use of `standard candles' such as the RC stars (cf. \citet{tatt13,choi18,gors20,skow21}), classical Cepheids \citep{niko04,inno16,deb18,josh19} to construct the reddening map of the LMC. The reddening values obtained in the present study are compared with those obtained by \citet{skow21}. For comparison, the $\Delta E(B-V)$ values are converted to the $\Delta E(V-I)$ values. Besides this, the reddening values $E(V-I)$ obtained from the \citet{skow21} reddening map are mean-subtracted using: $\overline{E(V-I)}= 0.100 \pm 0.043$. The residuals between the two sets of $\Delta E(V-I)$ values: one from this study and the other from \citet{skow21} are obtained. The histogram plot of residuals in Fig.~\ref{fig:fig5} shows that these values are normally distributed. However, a small offset of $\sim 0.03$ mag in the residuals is observed between each set of reddening values obtained from the PL relation considering break/no-break as compared to the \citet{skow21} map. Nonetheless, this shift lies within the error-bars of the average reddening value of the LMC obtained by \citet{skow21}. Hence the reddening values for the LMC Cepheids obtained in the present study are consistent with those obtained from the \citet{skow21} reddening map.

\begin{table*}
    \caption{Position angle of line of nodes $(\theta_{\rm lon})$ and inclination angle $(i)$ values of the LMC obtained in the present study.}
    \centering
    \begin{threeparttable}
    \begin{tabular}{c|c|c|c|c|c|c|}
        \hline
         \hline
         & & \multicolumn{2}{|c|}{${\rm Multi-phase}^{\rm a}$} & 
         \multicolumn{2}{|c|}{${\rm Mean ~Magnitude}^{\rm b}$} \\
         \hline
         & Cepheid Sample & Position Angle $(\theta_{\rm lon})$ & Inclination 
         Angle $(i)$ & Position Angle $(\theta_{\rm lon})$ & Inclination Angle $(i)$\\
         \hline
         \multirow{4}{*}{\shortstack{Without PL \\ Break}}& FU & $157\rlap{.}^{\circ}80^{\pm 1\rlap{.}^{\circ}70{\rm (stat.)}}_{\pm 1\rlap{.}^{\circ}01{\rm (syst.)}}$ & $24\rlap{.}^{\circ}06^{\pm 0\rlap{.}^{\circ}67{\rm (stat.)}}_{\pm 0\rlap{.}^{\circ}44{\rm (syst.)}}$ & 
         $157\rlap{.}^{\circ}34^{\pm 4\rlap{.}^{\circ}26 {\rm (stat.)}}_{\pm 1\rlap{.}^{\circ}01 {\rm (syst.)}}$ & $24\rlap{.}^{\circ}38^{\pm 1\rlap{.}^{\circ}68 {\rm (stat.)}}_{\pm 0\rlap{.}^{\circ}46 {\rm (syst.)}}$ \\
         & FO & $147\rlap{.}^{\circ}55^{\pm 1\rlap{.}^{\circ}57{\rm (stat.)}}_{\pm 1\rlap{.}^{\circ}01 {\rm (syst.)}}$ & $23\rlap{.}^{\circ}29^{\pm 0\rlap{.}^{\circ}56 {\rm (stat.)}}_{\pm 0\rlap{.}^{\circ}76 {\rm (syst.)}}$ &
         $147\rlap{.}^{\circ}65^{\pm 4\rlap{.}^{\circ}51 {\rm (stat.)}}_{\pm 1\rlap{.}^{\circ}01 {\rm (syst.)}}$ & $23\rlap{.}^{\circ}20^{\pm 1\rlap{.}^{\circ}58 {\rm (stat.)}}_{\pm 0\rlap{.}^{\circ}76 {\rm (syst.)}}$ \\
         & FU+FO & $152\rlap{.}^{\circ}45^{\pm 1\rlap{.}^{\circ}16 {\rm (stat.)}}_{\pm 1\rlap{.}^{\circ}01 {\rm (syst.)}}$ & $23\rlap{.}^{\circ}40^{\pm 0\rlap{.}^{\circ}43 {\rm (stat.)}}_{\pm 0\rlap{.}^{\circ}60 {\rm (syst.)}}$ & 
         $152\rlap{.}^{\circ}28^{\pm 3\rlap{.}^{\circ}16 {\rm (stat.)}}_{\pm 1\rlap{.}^{\circ}01 {\rm (syst.)}}$ & $23\rlap{.}^{\circ}70^{\pm 1\rlap{.}^{\circ}17 {\rm (stat.)}}_{\pm 0\rlap{.}^{\circ}60 {\rm (syst.)}}$ \\

         \hline
         \multirow{4}{*}{\shortstack{With PL \\ Break}}& FU & $159\rlap{.}^{\circ}44^{\pm 1\rlap{.}^{\circ}63 {\rm (stat.)}}_{\pm 1\rlap{.}^{\circ}01 {\rm (syst.)}}$ & $24\rlap{.}^{\circ}05^{\pm 0\rlap{.}^{\circ}65 {\rm (stat.)}}_{\pm 0\rlap{.}^{\circ}40 {\rm (syst.)}}$ &
         $157\rlap{.}^{\circ}98^{\pm 3\rlap{.}^{\circ}82 {\rm (stat.)}}_{\pm 1\rlap{.}^{\circ}01 {\rm (syst.)}}$ & $24\rlap{.}^{\circ}56^{\pm 1\rlap{.}^{\circ}56 {\rm (stat.)}}_{\pm 0\rlap{.}^{\circ}44 {\rm (syst.)}}$ \\
         & FO & $149\rlap{.}^{\circ}62^{\pm 1\rlap{.}^{\circ}63 {\rm (stat.)}}_{\pm 1\rlap{.}^{\circ}01 {\rm (syst.)}}$ & $22\rlap{.}^{\circ}30^{\pm 0\rlap{.}^{\circ}57 {\rm (stat.)}}_{\pm 0\rlap{.}^{\circ}69 {\rm (syst.)}}$ &
         $151\rlap{.}^{\circ}80^{\pm 4\rlap{.}^{\circ}62 {\rm (stat.)}}_{\pm 1\rlap{.}^{\circ}01 {\rm (syst.)}}$ & $22\rlap{.}^{\circ}98^{\pm 1\rlap{.}^{\circ}65 {\rm (stat.)}}_{\pm 0\rlap{.}^{\circ}62 {\rm (syst.)}}$ \\
         & FU+FO & $154\rlap{.}^{\circ}76^{\pm 1\rlap{.}^{\circ}16 {\rm (stat.)}}_{\pm 1\rlap{.}^{\circ}01 {\rm (syst.)}}$ & $22\rlap{.}^{\circ}87^{\pm 0\rlap{.}^{\circ}43 {\rm (stat.)}}_{\pm 0\rlap{.}^{\circ}53 {\rm (syst.)}}$ & 
         $155\rlap{.}^{\circ}52^{\pm 2\rlap{.}^{\circ}93 {\rm (stat.)}}_{\pm 1\rlap{.}^{\circ}01 {\rm (syst.)}}$ & $23\rlap{.}^{\circ}73^{\pm 1\rlap{.}^{\circ}13 {\rm (stat.)}}_{\pm {\rm 0\rlap{.}^{\circ}51 {\rm (syst.)}}}$ \\
         \hline
         \hline
    \end{tabular}
    \begin{tablenotes}
    % \item[a] Number of phase points in the multi-phase data.
    \item[a] The viewing angle parameters are obtained by using the multi-phase PL relations.
    \item[b] Mean magnitude PL relations are used to obtain the viewing angle parameters. Mean magnitudes are obtained from the flux-averaged values of the magnitudes corresponding to 50 phase points over a complete pulsation cycle. 
    \end{tablenotes}
    \end{threeparttable}
    \label{tab:1}
\end{table*}

\subsection{Geometry of the LMC}
\begin{figure}
    \centering
    \includegraphics[width=\linewidth, keepaspectratio, trim={0.1in 0.1in 0 0}, clip]{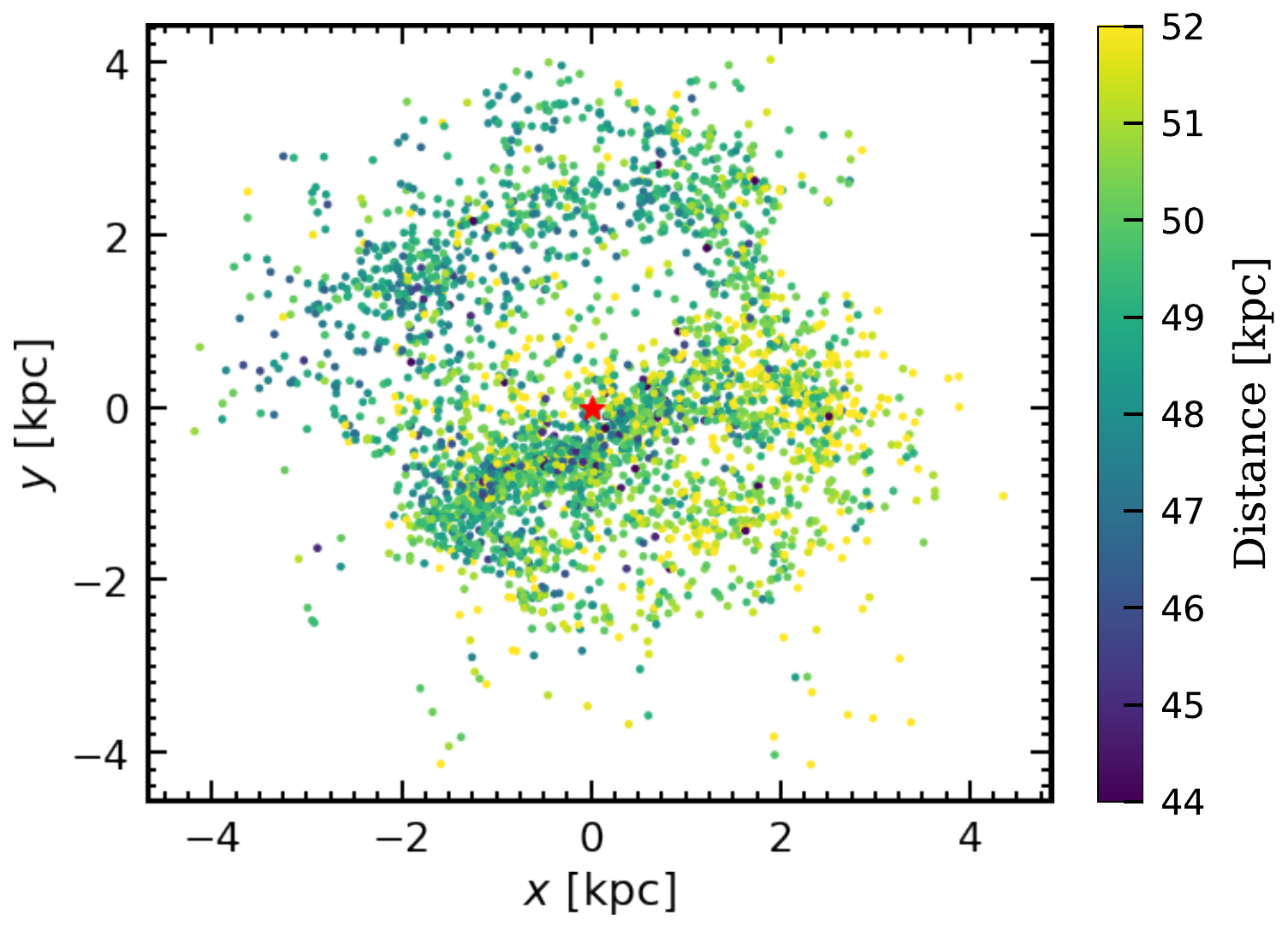}
    \caption{Distribution of $3293$ Cepheids in the $XY$ plane of the LMC based on absolute values of distances determined from the multi-phase PL relations considering a break in the present study. The Cartesian $(X,Y)$ coordinates are determined using the transformation relations in equation~\ref{eq:13}. The centre of the LMC is shown as a star symbol in red colour.}
    \label{fig:fig6}
\end{figure}

The determination of geometry of the LMC relies on distribution of Cepheids in three-dimensional Cartesian coordinates, $(x,y,z)$ which is determined using the equatorial coordinates $(\alpha_{i}, \delta_{i})$ and distance $D_{i}$. The distance modulus $(\mu_{0,i})$ values of Cepheids are converted to absolute distance values $(D_{i})$ in kiloparsecs (kpc) using the standard distance formula:
\begin{align}
D_{i} = & 10^{[0.2(\mu_{0,i} - 10)]}. \label{eq:18}   
\end{align}
The Cartesian $(x,y,z)$ values are obtained using the following transformation equations: \citep{wein01, mare01}
\begin{align}
x = & -D \sin{(\alpha - \alpha_{0})} \cos{\delta}, \nonumber \\ 
y = &D \sin{\delta}\cos{\delta_{0}} - D\sin{\delta_{0}} 
\cos{(\alpha - \alpha_{0})} \cos{\delta}, \\ \nonumber
z = & D_{0} - D \sin{\delta} \sin{\delta_{0}} - D \cos{\delta_{0}}\cos{(\alpha-\alpha_{0})} 
\cos{\delta}.
% \label{eq:19}
\end{align}
Here $(\alpha_{0}, \delta_{0})$ and $D_{0}$ denote the equatorial coordinates and distance to the centre of the LMC, respectively. The following values are adopted in the present study: $(\alpha_{0}, \delta_{0})$ = $(80.78, -69.03)$ \citep{niko04} and $D_{0} = 49.59$ kpc \citep{pie19}. The resulting $(x,y,z)$ coordinate system is such that the direction of $x-$axis is antiparallel to the $\alpha-$axis, the direction of $y-$axis is along the $\delta-$axis and the $z-$axis is directed towards the observer. The LMC has nearly a planar geometry \citep{mare01, inno16, deb18,ripe22}. The equation of the plane fitted to the distribution of Cepheids is mathematically given as \citep{niko04}:
\begin{align}
z = & Ax+By+C, \label{eq:20}
\end{align}
The values of the coefficients $A$, $B$ and $C$ are obtained from the best-fit solution of the plane equation. The values of angle of inclination $(i)$ and position angle of line of nodes $(\theta_{\rm lon})$ are determined using the following: 
\begin{align}
\theta_{\rm lon} =& \arctan{\left(-\frac{A}{B}\right)} + {\rm sign}(B) \frac{\pi}{2}, \nonumber \\
i =& \arccos{\left(\frac{1}{\sqrt{1+A^{2}+B^{2}}}\right)}. \label{eq:21}
\end{align}
The errors in the values of $i$ and $\theta_{\rm lon}$ are obtained from the propagation of errors in the coefficients $A$ and $B$. The equations used for estimating the errors in the values of $i$ and $\theta_{\rm lon}$ are given as follows:
\begin{align}
\sigma_{\theta} = & \frac{1}{\sqrt{A^{2}+B^{2}}} \sqrt{A^{2} \sigma_{B}^{2} + B^{2}
\sigma_{A}^{2}}, \nonumber \\
\sigma_{i} = & \left( \frac{1}{A^{2} + B^{2} + 1} \right) \frac{1}{\sqrt{A^{2}+B^{2}}} 
\sqrt{A^{2} \sigma_{A}^{2} + B^{2} \sigma_{B}^{2}}. \label{eq:22} 
\end{align}
Here, $\sigma_{A}$ and $\sigma_{B}$ represents the errors associated with the coefficients $A$ and $B$ respectively. The quantities $\sigma_{\theta}$ and $\sigma_{i}$ represents the corresponding uncertainties in position angle of line of nodes and inclination angle, respectively. The distribution of common Cepheids in the $XY$ plane of the LMC based on their absolute distance values determined from multi-phase PL relations considering break are shown in Fig.~\ref{fig:fig6}. 

Using multi-phase PL relations, the LMC viewing angles are determined using only FU- and  FO-mode Cepheids as well as the combined sample of (FU+FO)-mode Cepheids. The viewing angle parameters are also determined separately using PL relations with/without breaks. The viewing angle parameters and their uncertainties presented in Table \ref{tab:1} are obtained using the Monte Carlo simulations \citep{deb15}. The simulations are run for $10^{5}$ iterative steps, and in each iteration a plane-fit solution is obtained by randomizing the $(x,y,z)$ coordinates with their corresponding uncertainties $(\sigma_{x}, \sigma_{y}, \sigma_{z})$. Finally, the values of viewing angles and their statistical errors are obtained by carrying out a Gaussian-fit to the distribution of values of $i$ and $\theta_{\rm lon}$ obtained from the $10^{5}$ iterations. The final results are presented in Table~\ref{tab:1}.

In the present study, the mean-light PL relations are also obtained taking the average of multi-phase magnitudes over 50 pulsation phases. These mean-light PL relations are then used to determine the viewing angles of the LMC following the same methodology. The results for LMC viewing angles obtained based on mean-light PL relations are also presented in Table~\ref{tab:1}. A comparison between the results obtained based on multi-phase and mean-light PL relations shows that viewing angles are comparable within the quoted error bars. However, the statistical errors in the viewing angle parameters obtained based on multi-phase PL relations are found to be significantly smaller as compared to those using the mean-light PL relations.    

The values of position angle parameter $(\theta_{\rm lon})$ determined based on multi-phase PL relations with/without breaks are found to be: $154\rlap{.}^{\circ}76 \pm 1\rlap{.}^{\circ}16$ and, $152\rlap{.}^{\circ}{29} \pm 1\rlap{.}^{\circ}16$, respectively. Comparing with the values obtained in earlier studies as presented in Table \ref{tab:2}, position angles determined based on multi-phase PL relations with/without breaks are found to be within $0.1\sigma-1.4\sigma$ of those reported by \citet{niko04}, \citet{koer09} and \citet{deb18}, respectively. These values are also found to be consistent when compared with the results obtained based on mean-light PL relations in this study (Table \ref{tab:2}). However, the values of position angle differ by $2.5\sigma-6\sigma$ when compared with the values obtained by \citet{inno16} and \citet{ripe22}, respectively. These two studies make use of mean light optical and near-infrared (NIR) data, where the relative distances of individual Cepheids from the centre of the LMC were converted into absolute distances to determine the viewing angle parameters. The study of \citet{inno16} deals with the determination of the best estimates of viewing angle parameters calculated from the weighted average of the values obtained from various optical and NIR PW relations involving FO, FU and combined (FO+FU) Cepheids. These relations were obtained considering different adopted centres of LMC taken from the literature. On the other hand, in the study of \citet{ripe22}, the weighted average of the values obtained from PLK$_{s}$, PWJK$_{s}$ and PWVK$_s$ relations were taken as the best estimates for the viewing angle parameters using a single adopted centre. Furthermore, it is to be mentioned here that the different numbers of Cepheid samples were taken to establish the various PL and PW relations in the above two studies for the determination of viewing angle parameters. However, the present study utilizes simultaneous fitting of multi-phase PL relations in multi-wavelength photometric bands  considering a common sample of Cepheids taken from the OGLE-IV and Gaia DR3 databases to determine the viewing angle parameters. This might explain the possible discrepancies in the determination of viewing parameters between this study and the two aforementioned studies.

The values of inclination angles $(i)$ determined based on multi-phase PL relations with/without breaks are found to be: $22\rlap{.}^{\circ}87 \pm 0\rlap{.}^{\circ}43$ and, $23\rlap{.}^{\circ}40 \pm 0\rlap{.}^{\circ}43$, respectively. The values of inclination angles obtained in this study are found to be within $0.2\sigma-1.8\sigma$ levels of those obtained by \citet{koer09} and \citet{inno16}, respectively. However, there is a difference of $3\sigma-6\sigma$ between these values when compared with the other studies \citep{niko04,deb18,ripe22}. These differences in the inclination angle parameter might be attributed to using multi-phase/mean-light Cepheid PL relations in the present/earlier studies.

An independent estimate of the LMC viewing angle parameters using RC stars based on NIR data from the IRSF Magellanic Cloud Point-Source Catalogue gives the following values: $i = 23\rlap{.}^{\circ}5 \pm 0\rlap{.}^{\circ}4$ and $\theta_{\rm lon} = 154\rlap{.}^{\circ}5 \pm 1\rlap{.}^{\circ}2$ \citep{koer09}. Our results closely match with these values at the $(0.1-0.4)\sigma$ and $(0.3-1.7)\sigma$ levels considering non-linear/linear multi-phase PL relations, respectively. It is quite evident that the viewing angle parameter values show reduced errors while using multi-phase PL relations as compared to those obtained based on mean-light PL relations in the present study. On the other hand, the statistical errors in the viewing angle parameters are comparable to those obtained in literature using mean light PL relations \citep{inno16,deb18, ripe22} and RC stars \citep{koer09}.

\begin{table}
   \caption{Viewing angle parameters of the LMC based on Cepheid mean light PL relations \citep{niko04,inno16,deb18,ripe22} and RC stars \citep{koer09} in the literature. The values for the same without/with break in the PL relation as obtained in the present study are also given.}
    \centering
    \begin{tabular}{|c|c|c|c|}
        \hline
        \hline 
         Source & & $i$ & $\theta_{\rm lon}$ \\
        \hline
        \cite{niko04} & & $30\rlap{.}^{\circ}07 \pm 1\rlap{.}^{\circ}10$ & $151\rlap{.}^{\circ}0 \pm 2\rlap{.}^{\circ}4$ \\     
        \cite{koer09} & & $23\rlap{.}^{\circ}5 \pm 0\rlap{.}^{\circ}4$ & $154\rlap{.}^{\circ}5 \pm 1\rlap{.}^{\circ}2$ \\
        \cite{inno16} & & $25\rlap{.}^{\circ}05 \pm 1\rlap{.}^{\circ}15$ & $150\rlap{.}^{\circ}76 \pm 1\rlap{.}^{\circ}15$  \\
        \cite{deb18} & & $25\rlap{.}^{\circ}11 \pm 0\rlap{.}^{\circ}36 $ & 
        $154\rlap{.}^{\circ}70 \pm 1\rlap{.}^{\circ}38$ \\
        \cite{ripe22} & & $25\rlap{.}^{\circ}7 \pm 0\rlap{.}^{\circ}4 $ & $145\rlap{.}^{\circ}6 \pm 1\rlap{.}^{\circ}0$  \\
        \hline
        \multirow{4}{*}{\shortstack{This Work}} & \multirow{2}{*}{\shortstack{Without \\ PL break}} & $24\rlap{.}^{\circ}13^{\pm 1\rlap{.}^{\circ}18 {\rm (stat.)}}_{\pm 0\rlap{.}^{\circ}73 {\rm (syst.)}}$ & $148\rlap{.}^{\circ}46^{ \pm 3\rlap{.}^{\circ}27 {\rm (stat.)}}_{\pm 1\rlap{.}^{\circ}01 {\rm (syst.)}}$  \\
        & & & \\
        \cline{2-4}
        & \multirow{2}{*}{\shortstack{With \\ PL break}} & & \\
        & & $23\rlap{.}^{\circ}96^{ \pm 1\rlap{.}^{\circ}16 {\rm (stat.)}}_{\pm 0\rlap{.}^{\circ}53 {\rm (syst.)}}$ & $154\rlap{.}^{\circ}57^{\pm 3\rlap{.}^{\circ}02 {\rm (stat.)}}_{\pm 1\rlap{.}^{\circ}01 {\rm (syst.)}}$ \\

        \hline 
        \hline
    \end{tabular}
 
    \label{tab:2}
\end{table}

Most of the earlier studies on the LMC geometry have quoted only the statistical uncertainties in the viewing angle parameters. In addition to statistical uncertainties, we also attempt to quantify the systematic uncertainties in these parameters. The errors in the determination of viewing angle parameters are dominated by uncertainties in the individual distance measurements $D_{i}$ \citep{inno16}. However a significant contribution to the systematic uncertainty comes from mean distance and reddening of the LMC as well as PL zero-points. Adding these contributions in quadrature, we can write:
\begin{align}
\sigma_{\rm tot} = & \sqrt{\sigma_{D_{0}}^{2} + \frac{\left(\sigma_{\rm zp}^{2} + \sigma_{\overline{E(B-V)}}^{2}\right)D_{0}^{2}}{(2.1715)^{2}}}.
\label{eq:23}
\end{align}
Here $\sigma_{\rm zp}$ refers to the minimum zero-point uncertainty in the multi-phase multi-wavelength PL relations. Propagating these uncertainties, the systematic uncertainties in the measurement of $\theta_{\rm lon}$ and $i$ are calculated to be: $0\rlap{.}^{\circ}82$ and $0\rlap{.}^{\circ}43$, respectively. Further taking into account statistical uncertainties of 0.02 and 0.01 mag, respectively in the OGLE and Gaia band as systematic uncertainty in photometry \citep{sosz15,evan18}, the errors increase to $1\rlap{.}^{\circ}01$ and $0\rlap{.}^{\circ}53$, respectively. These error values are corresponding to the case of considering multi-phase PL relations with break to determine the LMC viewing angle parameters. We have calculated and presented both the statistical and systematic uncertainties in the viewing angle parameters for all the other cases in both Table~\ref{tab:1} and Table~\ref{tab:2}.   

\subsection{PC Relations at Maximum/Minimum Light and PLC relation at Mean Light}
The PC and PL relations of Cepheids are related to each other through the PLC relations. The PLC relations offer an insight in understanding the intrinsic scatter observed in the PC and PL relations \citep{kanb96}. In other words, the variation in surface temperatures of Cepheids corresponding to the same period is explained by the PLC relations. On the other hand, the hydrodynamics of the outer envelope of classical Cepheids can be extensively studied and understood using the PC relations \citep{skm93,kanb04,bhar14,das20}. 

In this study, the nature of PC and PLC relations of the FU-mode Cepheids are explored at phases corresponding to extinction corrected maximum, minimum, and mean light based on different reddening values. The reddening values obtained from the multi-phase PL relations with/without break are used for this purpose. The resulting PC relations are then compared to the one where the extinction correction is done based on the \citet{skow21} reddening map. The $(V-I)$ colour terms corresponding to the OGLE ($V,I$) bands are determined using maximum, minimum, and mean light as in the following \citep{kanb04, kanb06}:
\begin{align}
(V-I)_{\rm max} =& V_{\rm max} - I_{\rm phmax}, \nonumber \\
(V-I)_{\rm min} =& V_{\rm min} - I_{\rm phmin}, \nonumber \\
(V-I)_{\rm mean}=& \langle V \rangle - \langle I \rangle. \label{eq:28}
\end{align}
Here $I_{\rm phmax}$/$I_{\rm phmin}$ represents the $I-$band magnitude corresponding to the phase of maximum/minimum light in the $V$ band. The terms $\langle I \rangle $ and $\langle V \rangle$ denote the mean light in $V$ and $I$ band, respectively. 

The existence of breaks in the PC relations of FU-mode Cepheids in the LMC at a period $10$ d have been reported in several earlier studies \citep{kanb04,bhar14,das20,kurb23}. Hence, the PC relations are fitted with/without considering PC breaks at $P=10$ d. The resulting fitted values are shown in Table \ref{tab:3}. The parameters $\alpha$, $\beta$ denote the intercept and slope of the PC relations, respectively. The subscripts \textit{all}, $S$ and $L$ denote models covering all, short and long-period Cepheids, respectively.

The PC slopes obtained at maximum magnitude without considering break are found to be similar in either of the cases where the extinction corrections are implemented using the derived reddening values and those obtained from \citet{skow21}. The same is also found to be true for the PC slopes obtained at minimum light. Furthermore, the slopes of PC relations at minimum light are found to be steeper than at maximum light, and is consistent with the results obtained in earlier studies \citep{kanb04,bhar14,kurb23}.

\begin{table*}
    \caption{PC relations of FU-mode Cepheids in the LMC used in this study.}
    \centering
    \begin{threeparttable}
    \begin{tabular}{c|c|c|c|c|c|c|c|c|c}
    \hline
    \hline
    $E(B-V)$ source & $\alpha_{\rm (all)}^{\rm a}$ & $\beta_{\rm (all)}^{\rm b}$ & $\sigma_{\rm (all)}$ & $\alpha_{\rm L}$ & $\beta_{\rm L}$ & $\sigma_{\rm L}$ & $\alpha_{\rm S}$ & $\beta_{\rm S}$ & $\sigma_{\rm S}$ \\
    \hline
    \multicolumn{10}{|c|}{Max} \\  
    \shortstack{$\rm{MPPL}^{\rm c}$ with break} & $0.237 \pm 0.005$ & $0.280 \pm 0.008$ & $0.07$ & $0.62 \pm 0.06$ & $-0.09 \pm 0.05$ & $0.07$ & $0.196 \pm 0.005$ & $0.357 \pm 0.009$ & $0.06$ \\
    \shortstack{MPPL without break} & $0.237 \pm 0.005$ & $0.281 \pm 0.007$ & $0.07$ & $0.80 \pm 0.06$ & $-0.26 \pm 0.05$ & $0.07$ & $0.179 \pm 0.005$ & $0.388 \pm 0.009$ & $0.06$ \\
    \cite{skow21} & $0.288 \pm 0.008$ & $0.287 \pm 0.012$ & $0.11$ & $0.66 \pm 0.09$ & $-0.08 \pm 0.08$ & $0.11$ & $0.242 \pm 0.009$ & $0.366 \pm 0.014$ & $0.10$  \\
    \multicolumn{10}{|c|}{Min} \\
    \shortstack{MPPL with break} & $0.544 \pm 0.003$ & $0.304 \pm 0.005$ & $0.05$ & $0.34 \pm 0.05$ & $0.51 \pm 0.04$ & $0.06$ & $0.580 \pm 0.003$ & $0.234 \pm 0.005$ & $0.04$ \\
    \shortstack{MPPL without break} & $0.544 \pm 0.003$ & $0.305 \pm 0.005$ & $0.05$ & $0.56 \pm 0.05$ & $0.30 \pm 0.04$ & $0.06$ & $0.582 \pm 0.003$ & $0.229 \pm 0.005$ & $0.04$ \\
    \cite{skow21} & $0.595 \pm 0.006$ & $0.311 \pm 0.009$ & $0.09$ & $0.36 \pm 0.08$ & $0.54 \pm 0.07$ & $0.09$ & $0.623 \pm 0.005$ & $0.248 \pm 0.008$ & $0.06$ \\
    \hline 
    \hline
    \end{tabular}
    \begin{tablenotes}
    \item[a] $\alpha$ parameters denote the intercepts of the PC relations.
    \item[b] $\beta$ parameters denote the slopes of the PC relations.
    \item[c] MPPL is the acronym for multi-phase PL relations.  
    \end{tablenotes}
    \end{threeparttable}
    \label{tab:3}
\end{table*}

\begin{figure*}
    \centering
    \includegraphics[width=1.0\textwidth, keepaspectratio]{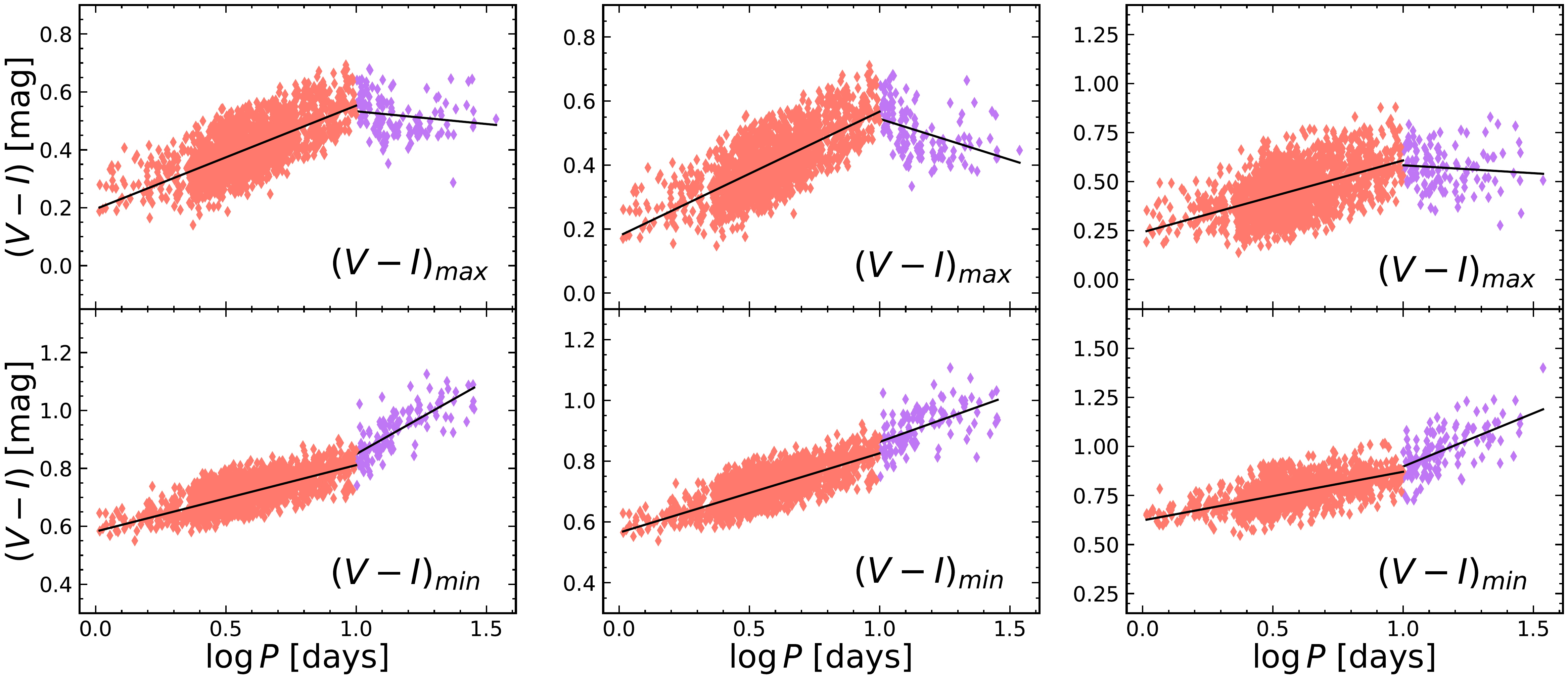}
    \caption{PC relations of  LMC FU mode Cepheids with breaks at $P = 10$ days. The figures in the first two panels show the PC relations at maximum/minimum light where the magnitudes are corrected for extinction based on the reddening values obtained from
    non-linear and linear multi-phase PL relations, respectively. The figures in the third panel depict the same obtained using  \citet{skow21} reddening map.}
    \label{fig:fig7}
\end{figure*}

The results of extinction-corrected PC relations considering a break at $P=10$ d are obtained based on three different sets of reddening values and are plotted in Fig.~\ref{fig:fig7}. The figures in the first two panels depict the PC relations when maximum/minimum light are extinction corrected using the reddening values obtained from multi-phase PL relations with/without break, respectively. On the other hand, the figures in the third panel represent the extinction-corrected PC relations at maximum/minimum light based on the reddening values obtained from the \citep{skow21} reddening map. Although the nature of the PC slopes and intercepts for each column show expected behaviour at maximum/minimum light, slight variations in the steepness of the PC slopes/intercepts may be observed.

In particular, the long period FU-mode Cepheid PC relations at maximum light are found to be flatter as compared to short period ones in all the cases. However, utilizing the reddening values derived from multi-phase PL relations without break yields a slightly sloped PC relation as compared to the other two sets. On the other hand, the slopes of extinction corrected PC relations at minimum light are found to be slightly more for long period Cepheids when compared with the short period counterparts, which is consistent with the results in the literature. The change in slopes of PC relations at minimum light in the long period range are found to be more using the reddening values obtained from both the multi-phase PL relations with break and the \citet{skow21} reddening map (refer to the first and third panels in Fig. \ref{fig:fig7}). However, only a marginal change can be observed in PC slopes for the long period Cepheids with respect to the short period ones (refer to the second column in Fig.~\ref{fig:fig7} when the reddening values obtained from multi-phase PL relations with no-break are considered.

Furthermore, the slopes of PC relations for long period FU Cepheids using minimum light are found to be steeper as compared to those using maximum light, independent of the choice of reddening map. This is an important result. The flatter PC relations at the phase of maximum light as compared to those at minimum light well captures the interaction of the HIF with the stellar photosphere of Cepheids. This has been reported in details by several earlier studies, both empirically and theoretically \citep{kanb04,bhar14,das20,kurb23}. Thus, it can be argued that the application of the reddening values obtained in this study yields results that are consistent with the physics of stellar pulsation for Cepheids.    
\begin{table*}
    \centering
    \caption{Mean magnitude PLC relations of FU mode Cepheids in the LMC with the $F$-test values for the \textit{colour} term.}
    \begin{threeparttable}
    \begin{tabular}{c|c|c|c|c}
    \hline
    \hline
    $E(B-V)$ source & $\alpha_{\lambda}^{\rm a}$ &  \multicolumn{2}{|c|}{Slopes} & $F$-Value \\ 
    & & $\beta_{\lambda}^{\rm b}$ & $C_{\lambda}^{\rm c}$ & $(\times 10^{3})$ \\
    \hline
    \multicolumn{5}{|c|}{$V$-band} \\ 
    \shortstack{$\rm{MPPL}^{\rm d}$ with break} & $15.90 \pm 0.02$ & $-3.26 \pm 0.02$ & $2.43 \pm 0.02$ & $18.10$  \\
    \shortstack{$\rm{MPPL}$ without break} & $15.90 \pm 0.02$ & $-3.26 \pm 0.02$ & $2.43 \pm 0.02$ & $18.21$  \\
    \citet{skow21} & $15.96 \pm 0.02$ & $-3.24 \pm 0.02$ & $2.34 \pm 0.02$ & $11.45$  \\
    \multicolumn{5}{|c|}{$I$-band} \\
    \shortstack{$\rm{MPPL}$ with break} & $15.84 \pm 0.01$ & $-3.19 \pm 0.01$ & $1.47 \pm 0.01$ & $13.25$ \\
    \shortstack{$\rm{MPPL}$ without break} & $15.84 \pm 0.01$ & $-3.19 \pm 0.01$ & $1.48 \pm 0.01$ & $16.10$  \\
    \citet{skow21} & $15.88 \pm 0.01$ & $-3.17 \pm 0.01$ & $1.42 \pm 0.01$ & $5.15$ \\
    \hline
    \hline
    \end{tabular}
    \begin{tablenotes}
    \item[a] $\alpha$ parameters denote the intercepts of the PLC relations.
    \item[b] $\beta$ parameters denote the coefficients of $\log{P}$ in the PLC relations.
    \item[c] $C$ parameters denote the coefficients of $(V-I)_{\rm mean}$ in the PLC relations.
    \item[d] MPPL is the acronym for multi-phase PL relations.  
    \end{tablenotes}
    \end{threeparttable}
    \label{tab:4}
\end{table*}    

The results of PLC relations using mean light in OGLE $V$ and $I$ bands are presented in Table \ref{tab:4}. The PLC relations are obtained separately for three sets in both the photometric bands using the mean light corrected for  extinction using the three reddening maps. The results show that the coefficient of the colour term $(V-I)$ is a non-zero positive number in each case. The results of the statistical $F-$ test for the existence of the colour term are also presented in Table \ref{tab:4}. The $p(F)$ values corresponding to the $F-$values of the PLC relations are all found to be $< 10^{-5}$, indicating a significant $(V-I)$ colour term at mean light. Furthermore, these results show that the PLC relations are consistent with each other independent of the choice of reddening map.          
\section{Summary And Conclusions} \label{summ}
The archival light curve data of more than $3290$ common Classical Cepheids in the LMC available in OGLE-IV ($V$, $I$) and Gaia photometric bands $(G_{\rm BP}$, $G$ and $G_{\rm RP})$ are exploited for the analysis in the present study. The multi-band Cepheid light curves are phased and multi-phase data are extracted for $50$ phase points over a complete pulsation cycle. 

The multi-phase PL relations are obtained in all the five bands by applying a two-step iterative method. In the first step, multi-phase PL relations are obtained without correcting for distance modulus and reddening. The PL relations are obtained with/without considering PL break at $P=10$ days for FU-mode and at $P=2.5$ days for FO-mode Cepheids. These multi-phase PL relations are used to determine the apparent distance modulus values $(\Delta \mu_{i,j})$ of individual Cepheids in multi-wavelength bands. The individual values of true distance modulus $(\Delta \mu_{0,i})$ and reddening $(\Delta E(V-I)_{i})$ of Cepheids relative to the mean LMC values are obtained in the second step by carrying out a simultaneous fit to the apparent distance modulus values with a reddening law using $R_{V} = 3.23$. 

The absolute values of distance modulus and reddening are determined for individual Cepheids by adding the values of $\Delta \mu_{0,i}$ and $\Delta E(V-I)_{i}$ to the mean LMC values taken from the literature. The  values of distance ($D$) along with the given RA-DEC $(\alpha, \delta)$ values are then used to determine the three-dimensional distribution of Cepheids in the LMC in terms of the Cartesian $(x, y, z)$ coordinates. The reddening maps are constructed using PL relations with/without break. The maps are found to be consistent with each other as well as with those obtained in the literature.

The LMC viewing angle parameters are obtained by fitting a plane of the form $z=f(x,y)$ to the three-dimensional distribution of Cepheids in the disk of LMC in Cartesian coordinates $(x, y, z)$. The inclination $(i)$ and position angles $(\theta_{\rm lon})$ are determined based on PL relations with/without break. The values of position angles are found to be within $0.1\sigma-1.4\sigma$ of those obtained by \citet{niko04}, \citet{koer09} and \citet{deb18}, respectively. However, position angle parameter values obtained in this study have been found to differ by $2.5\sigma-6\sigma$ as compared to those obtained by \citet{inno16} and \cite{ripe22}. Furthermore, the inclination angle parameters obtained in this study are found to be consistent with that obtained by \citet{koer09} and \citet{inno16} within $0.1\sigma - 1.4\sigma$ levels. However, there is a difference of $3\sigma-6\sigma$ in the inclination angle parameters as compared to those obtained in some earlier studies \citep{niko04,deb18,ripe22}. The results of LMC viewing angle parameters closely match with the values as those obtained by \citet{koer09} using RC stars. The systematic uncertainty in the determination of LMC viewing angle parameters is found to be $0\rlap{.}^{\circ}82$ and $0\rlap{.}^{\circ}43$, respectively. These increase to $1\rlap{.}^{\circ}01$ and $0\rlap{.}^{\circ}53$, respectively when the systematic uncertainties in the OGLE and Gaia band photometry are propagated.  

The use of multi-phase PL relations significantly reduces the statistical and systematic uncertainties in the viewing angle parameters as compared to those obtained from the mean-light PL relations in this study. This demonstrates a clear advantage of using multi-phase PL relations over the mean-light PL relations in determining the LMC geometry with improved accuracy and precision.

% \textbf{We have investigated the existence of possible warped and flared structures in the LMC disk. Results obtained in the present study suggest the presence of an asymmetric warp in the disk of LMC. We have also found evidence for a flared structure in the LMC disk. The disk has varying scale height which increases rapidly in radially outward direction.}
The PC relations of the FU-mode Cepheids in the LMC are obtained using maximum and minimum light in OGLE $V$ and $I$ bands with/without considering break at $P=10$ d. The PC relations are obtained separately by applying extinction correction using the reddening values obtained in this study with/without PL break and those from the \cite{skow21} reddening map. The PC relations at maximum/minimum light without considering break are found to be free from the choice of reddening values used. On the other hand, when a break $P=10$ d is considered, the PC relations are found to be weakly sensitive to the choice of reddening values. However, the PC slopes at maximum light are found to be flatter than those at minimum light and is independent of the choice of reddening values. These results thus show that the extinction-corrected PC relations based on the reddening values determined from the multi-phase PL relations are consistent with the findings of earlier studies and can accurately describe the physics of stellar pulsation in Cepheids. 

The PLC relations of the FU mode Cepheids using mean light are obtained in OGLE ($V,I$)
photometric bands. The PLC relations are obtained in three sets by applying extinction correction to mean light using the reddening values obtained from multi-phase PL relations with/without break and those from \citet{skow21}. The PLC relations are found to be similar in each band for all the three sets. The result of statistical $F-$test shows that the $(V-I)$ colour term is highly significant in both bands.    

\section*{Acknowledgements}
The authors acknowledge the use of highly valuable publicly accessible archival data from OGLE-IV and Gaia DR3. GB is grateful to the Department of Science and Technology (DST), Govt. of India, New Delhi for providing the financial support of this study as a Junior Research Fellow (JRF) through the DST INSPIRE Fellowship research grant (DST/INSPIRE/Fellowship/2019/IF190616). SD thanks CSIR, Govt. of India, New Delhi for the financial support received through a research grant ``03(1425)/18/EMR-II''. MD thanks CSIR, Govt. of India, New Delhi for the JRF provided through CSIR-NET under the project. SMK thanks State University of New York, Oswego, NY 13126, USA and Cotton University, Guwahati, Assam for the support. AB acknowledges funding from the European Union’s Horizon 2020 research and innovation programme under the Marie Skłodowska-Curie grant agreement No. 886298. Funding for the Stellar Astrophysics Centre was provided by The Danish National Research Foundation (Grant DNRF106). The authors acknowledge IUCAA, Pune for providing access to the Pegasus High Performance Computing facility. Finally, the authors are grateful to the anonymous referee for useful comments and constructive suggestions which have significantly improved the presentation of the manuscript.   
% %%%%%%%%%%%%%%%%%%%%%%%%%%%%%%%%%%%%%%%%%%%%%%%%%%
\section*{Data Availability}
The OGLE-IV data is collected from \url{http://ftp.astrouw.edu.pl/ogle/ogle4/OCVS/lmc/cep/}. The Gaia DR3 data is downloaded from \url{http://cdn.gea.esac.esa.int/Gaia/gdr3/} using Python astro-query. Cross-matching with the OGLE data is done using CDS-Xmatch(\url{http://cdsxmatch.u-strasbg.fr/}) service. The \citet{skow21} reddening map is downloaded from \url{http://ogle.astrouw.edu.pl/}.

% The inclusion of a Data Availability Statement is a requirement for articles published in MNRAS. Data Availability Statements provide a standardized format for readers to understand the availability of data underlying the research results described in the article. The statement may refer to original data generated in the course of the study or to third-party data analysed in the article. The statement should describe and provide means of access, where possible, by linking to the data or providing the required accession numbers for the relevant databases or DOIs.

%%%%%%%%%%%%%%%%%%%% REFERENCES %%%%%%%%%%%%%%%%%%

% The best way to enter references is to use BibTeX:

\bibliographystyle{mnras}
\bibliography{example} % if your bibtex file is called example.bib

%%%%%%%%%%%%%%%%%%%%%%%%%%%%%%%%%%%%%%%%%%%%%%%%%%

%%%%%%%%%%%%%%%%% APPENDICES %%%%%%%%%%%%%%%%%%%%%

\appendix
\section{Additional Figures}
\begin{figure*}
    \centering
     \includegraphics[width=0.9\textwidth, keepaspectratio]{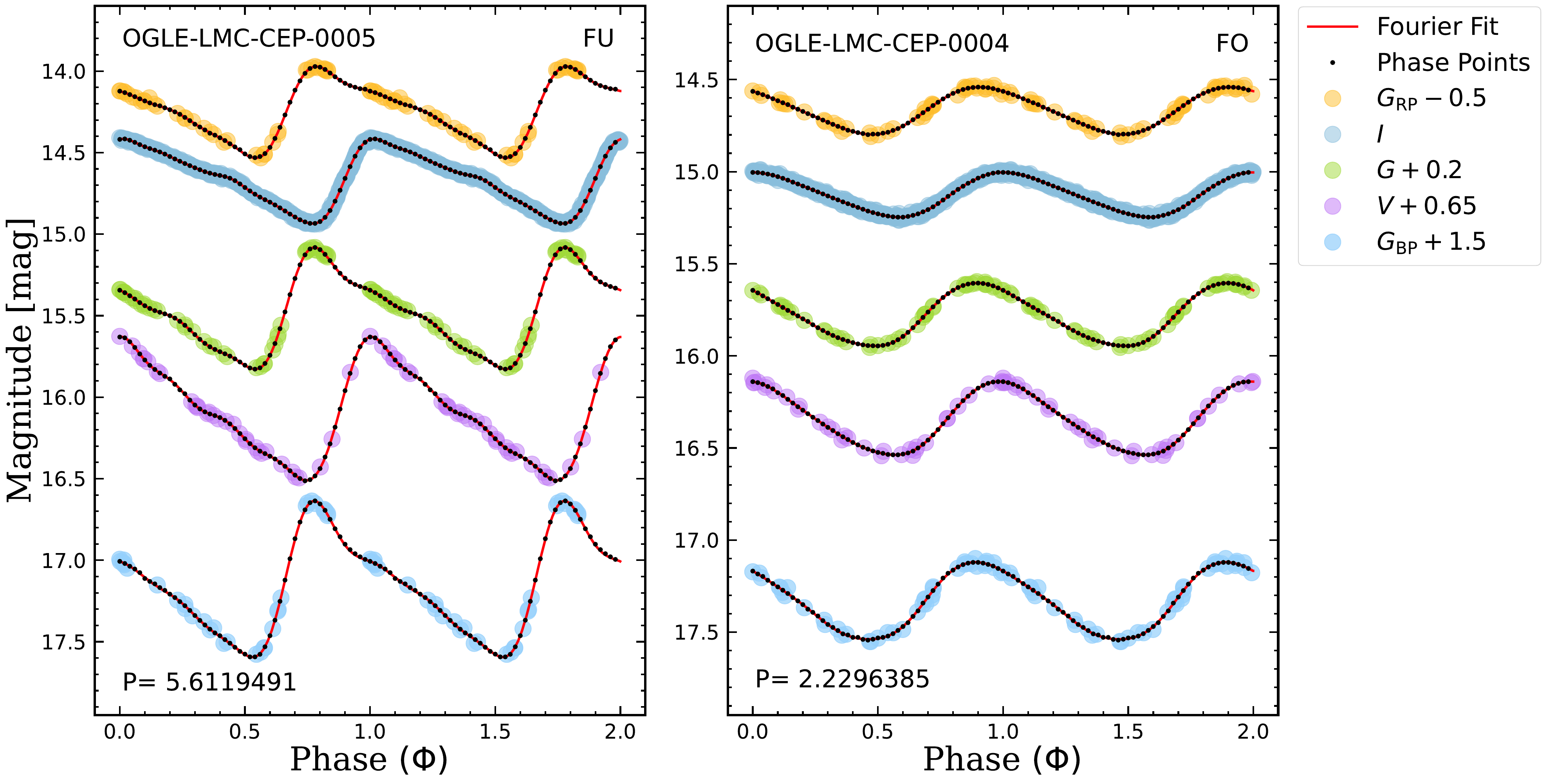}
    \caption{Representative light curves of an FU and FO mode Cepheid in the LMC in OGLE $(V,I)$ and Gaia $(G,G_{\rm BP},G_{\rm RP})$ photometric bands. The light curves are shifted in phase with respect to the epoch of maximum brightness in the $V$ band. The Fourier fit is shown with a solid line in red colour and the magnitudes extracted at 50 different phase points on the Fourier fitted line are over-plotted with black points.}
    \label{fig:a0}
\end{figure*}

\begin{figure*}
    \centering
    \includegraphics[width=0.9\textwidth, keepaspectratio]{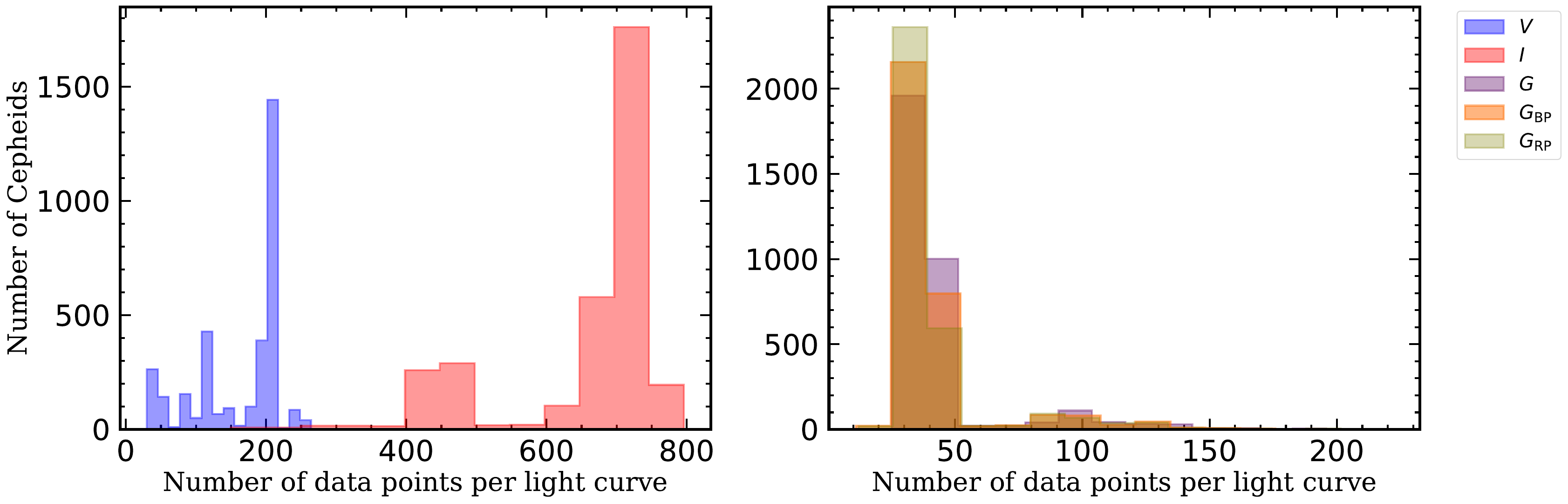}
     \caption{Distribution of the number of photometric measurements in the light curves of LMC Cepheids in in OGLE $(V,I)$ and Gaia $(G,G_{\rm BP},G_{\rm RP})$ bands.}
     \label{fig:a01}
\end{figure*}
\begin{figure*}
     \centering
     \includegraphics[width=0.9\textwidth, keepaspectratio]{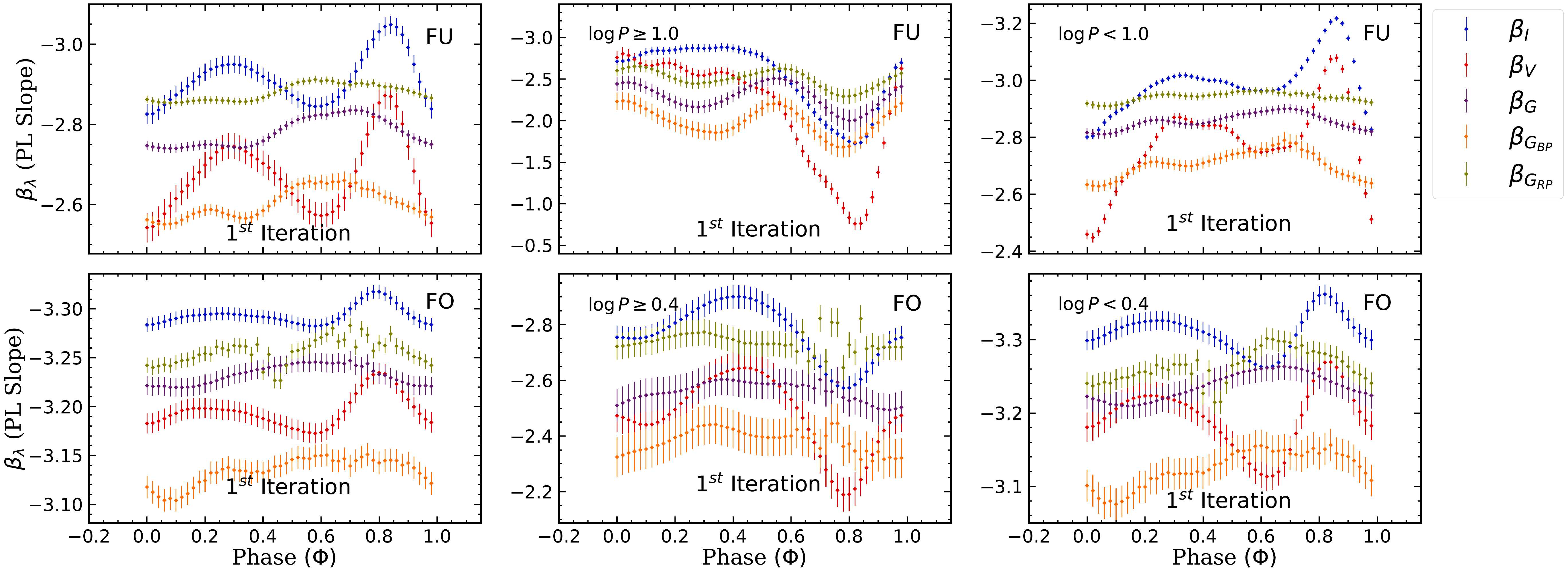}
     \caption{Same as Fig.\ref{fig:fig2} obtained from the first iteration.}
     \label{fig:a1}
\end{figure*}

\begin{figure*}
    \centering
    \includegraphics[width=0.9\textwidth, keepaspectratio]{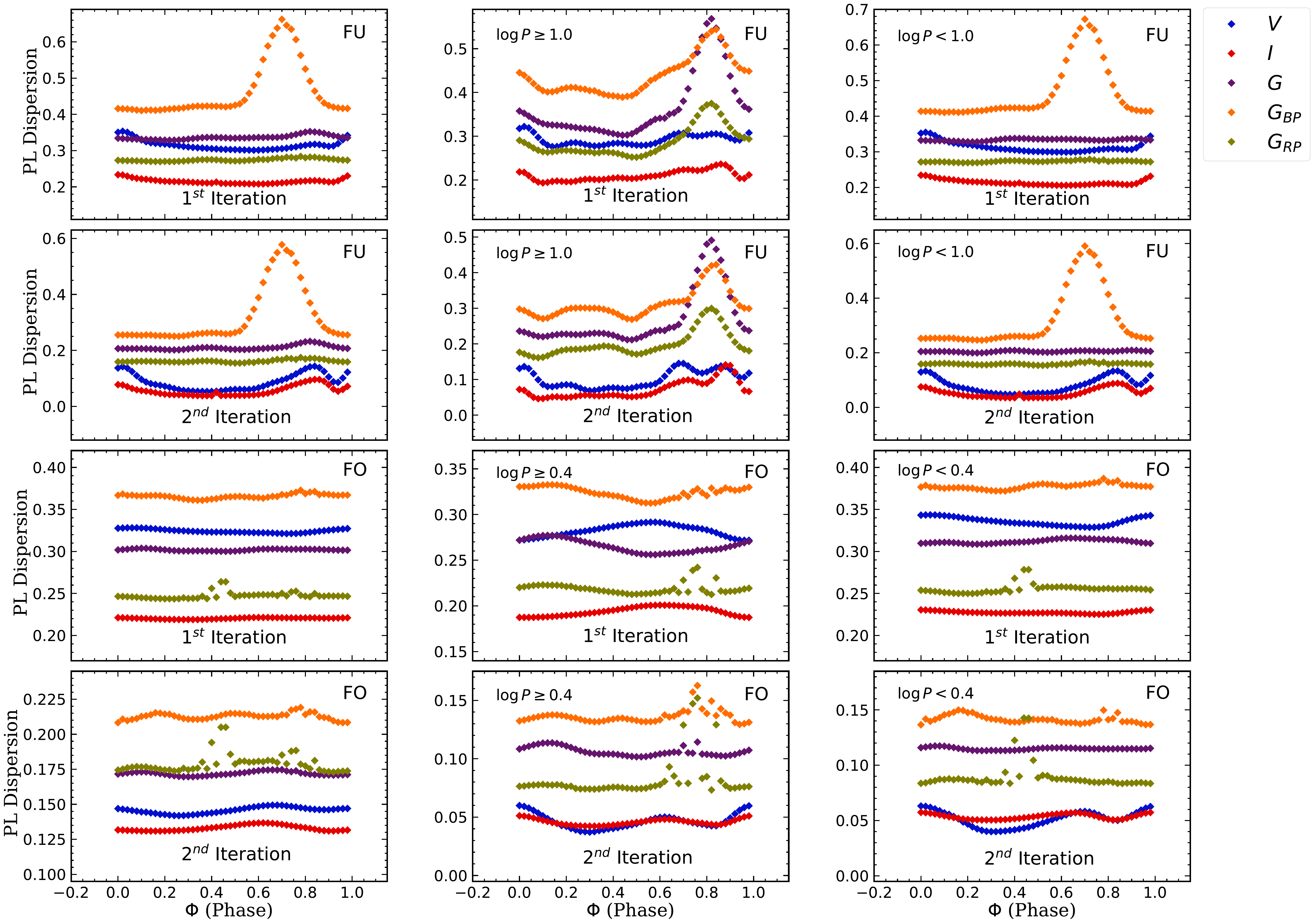}
    \caption{Dispersion in multi-phase PL relations of multi-photometric ($V,I,G,G_{\rm BP},G_{\rm RP}$) bands as a function of pulsation phase $(\Phi)$ for both FU \& FO-mode Cepheids before (first iteration) and after (second iteration) distance and reddening corrections, respectively. The figures in the \textit{left} panel show the dispersion in multi-phase PL relations for both FU and FO-mode Cepheids when no PL break is considered. The figures in the \textit{middle} and \textit{right} panels demonstrate the same considering a PL break for both long and short period FU \& FO-mode Cepheids, respectively.}
    \label{fig:a2}
\end{figure*}

\begin{figure*}
    \centering
    \includegraphics[width=0.9\textwidth, keepaspectratio]{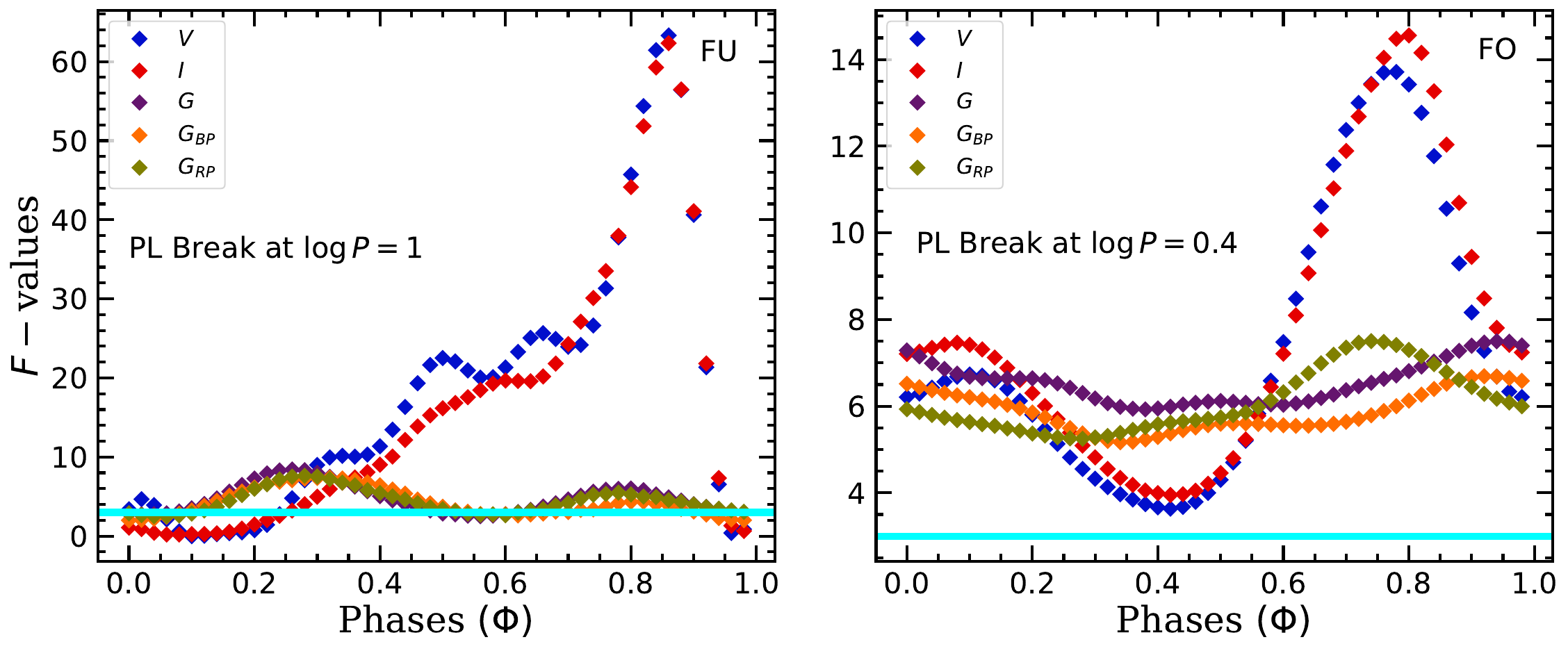}
    \caption{Comparison of $F$-values for the multi-phase PL relations of FU and FO-mode Cepheids in the LMC obtained using $100$ phase points over a complete pulsation cycle for OGLE-IV and Gaia bands. The \textit{cyan} coloured horizontal line corresponds to $F \sim 3.0$ which represents the $95\%$ confidence level (or p(F)=0.05). $F-$values above this line imply the presence of non-linearity in the PL relation.}
    \label{fig:a3}
\end{figure*}
\newpage
\section{System of equations in matrix form}\label{appendix}
The multi-phase PL relations, distance modulus and the reddening values are obtained using equations~\eqref{eq:5}-\eqref{eq:10} as described in Section \ref{method2}. The matrix equations presented in the Section \ref{method2} are the generalized formulations of equations employed to obtain the parameters. However, in practice the two-step iterative method makes use of different sets of matrices in the matrix equations to solve for the multi-phase PL relations, distance modulus and the reddening values. Two different types of GLM are used to obtain the multi-phase PL relations: one without considering PL break or the ``half model'' and the other allowing for a PL break or the ``full model''. The equations presented  here in matrix form are given in terms of the full model. The generalized linear model used to obtain the multi-phase PL relations in the first step is of the form:
\begin{align}
\mathbf{y_{1}} =& \mathbf{X_{1} q_{1}} + \xi_{1}. \label{eq:b1}    
\end{align}
Here the vector $\mathbf{y_{1}}$ contains the magnitudes $(m_{\lambda, i, j})$ of Cepheids in multiple wavelength at multiple phases. Here $i=1,2,3 \hdots, N_{\rm t}$; where $N_{\rm t}$ denotes the total number of stars. Furthermore, $j=1,2,3, \hdots, J$,  where $J=50$ represents the total number of phase points. The vector $\mathbf{q_{1}}$ denotes the parameter matrix containing the set of multi-phase PL slope $(\beta_{\lambda, j})$ and intercept $(\alpha_{\lambda, j})$ parameters in all photometric bands. The matrices $\mathbf{y}_{1}$ and $q_{1}$ can be expressed in the following form:
\begin{align}
\mathbf{y}_{1} = &  \begin{bmatrix} m_{V,1,1}^{l} \\  m_{V,2,1}^{l} \\  m_{V,3,1}^{l} \\ \vdots \\  m_{V,N_{1},1}^{l} \\  m_{V,1,1}^{s} \\ \vdots \\ m_{V,N_{2},1}^{s} \\ \vdots \\  m_{V,N_{2},J}^{s} \\ \vdots \\ m_{G_{{\rm RP}}, N_{2},J}^{s} \end{bmatrix}_{(N_{\rm t}\times J\times 5, 1)};
\hspace{0.5cm}
\mathbf{q_{1}} = \begin{bmatrix} \alpha_{V,1}^{l} \\ \beta_{V,1}^{l} \\ \alpha_{V,1}^{s} \\ \beta_{V,1}^{s} \\ \vdots \\ \alpha_{V, J}^{l} \\ \beta_{V,J}^{l} \\ \alpha_{V, J}^{s} \\ \beta_{V,J}^{s} \\ \vdots \\ \alpha_{G_{\rm RP}, J}^{l} \\ \beta_{G_{\rm RP},J}^{l} \\ \alpha_{G_{\rm RP}, J}^{s} \\ \beta_{G_{\rm RP},J}^{s} \end{bmatrix}_{(4J \times 5 ,1)}. 
\label{eq:b2}
\end{align}
Here the superscripts $l$ and $s$ stand for long and short period Cepheids. The number of long and short period Cepheids are denoted by $N_{1}$ and $N_{2}$, respectively, and $N_{\rm t}=N_{1} + N_{2}$. The matrix $\mathbf{X_{1}}$ contains the set of equations representing the multi-phase PL relations in all the five photometric bands. It is also referred to as the design matrix or the equation matrix. It is given in the form:
\clearpage
\begin{align}
\mathbf{X_{1}} = & \begin{bmatrix}
1 & \log{P_{1}^{l}} & 0 & 0 & \hdots & 0 & 0\\
\\ 
1 & \log{P_{2}^{l}} & 0 & 0 & \hdots & 0 & 0\\
\vdots  & \vdots & \vdots & \vdots & \hdots & \vdots & \vdots \\
1 & \log{P_{N_{1}}^{l}} & 0 & 0 & \hdots & 0 & 0\\ 
\\
0 & 0 &  1 & \log{P_{1}^{s}} & \hdots & 0 & 0\\
\\
0 & 0 &  1 & \log{P_{2}^{s}} & \hdots & 0 & 0\\
\vdots  & \vdots & \vdots & \vdots & \hdots & \vdots & \vdots \\
0 & 0 &  1 & \log{P_{N_{2}}^{s}} & \hdots & 0 & 0\\
\vdots  & \vdots & \vdots & \vdots & \ddots & \vdots & \vdots \\
0 & 0 & 0 & 0 & \hdots & 1 & \log{P_{N_{2}}^{s}}
\end{bmatrix}_{(N_{\rm t}\times J\times 5, ~4J\times 5)}.
\label{eq:b3}
\end{align}
The dimension or order of the matrices in the above are given as their subscripts in parentheses. The system of equations represented by equation~\eqref{eq:10} to obtain the corrections to true reddening values and distance moduli of individual Cepheids are represented by the GLM of the form:
\begin{align}
\mathbf{y_{2}} =& \mathbf{X_{2} q_{2}} + \xi_{2}. \label{eq:b4}    
\end{align}
Here the apparent distance modulus values of Cepheids determined based on multi-phase PL relations constitute the vector $\mathbf{y_{2}}$. The vector $\mathbf{q_{2}}$ represents the parameter matrix. In matrix form, they can be expressed as follows:
\begin{align}
\mathbf{y}_{2} = & \begin{bmatrix} \Delta\mu_{V,1,1}^{l} \\ \Delta\mu_{V,2,1}^{l} \\ \Delta\mu_{V,3,1}^{l} \\ \vdots \\ \Delta\mu_{V,N_{1},1}^{l} \\ \Delta\mu_{V,1,1}^{s} \\ \vdots \\ \Delta\mu_{V,N_{2},1}^{s} \\ \vdots \\ \Delta\mu_{V,N_{2},J}^{s} \\ \vdots \\ \Delta\mu_{G_{{\rm RP}}, N_{2},J}^{s} \end{bmatrix}_{(N_{\rm t}\times J\times 5, ~1)};
\hspace{0.5cm}
\mathbf{q_{2}} = \begin{bmatrix} \Delta\mu_{0,1}^{l} \\ \Delta E(B-V)_{0,1}^{l} \\ \Delta\mu_{0,2}^{l} \\ \Delta E(B-V)_{0,2}^{l} \\ \vdots \\ \Delta\mu_{0,N_{1}}^{l} \\ \Delta E(B-V)_{0,N_{1}}^{l} \\ \Delta\mu_{0,1}^{s} \\ \Delta E(B-V)_{0,1}^{s} \\ \vdots \\ \Delta\mu_{0,N_{2}}^{s} \\ \Delta E(B-V)_{0,N_{2}}^{s} \end{bmatrix}_{(2N_{\rm t},~1)}. \label{eq:b5}
\end{align}
The vector $\mathbf{X_{2}}$ has the form:
\begin{align}
\mathbf{X_{2}} = & \begin{bmatrix}
1 & R_{V,1} & 0 & 0 & \hdots & 0 & 0\\
1 & R_{V,2} & 0 & 0 & \hdots & 0 & 0\\
\vdots  & \vdots & \vdots & \vdots & \hdots & \vdots & \vdots \\
1 & R_{V, N} & 0 & 0 & \hdots & 0 & 0\\ 
0 & 0 &  1 & R_{V,1} & \hdots & 0 & 0\\
0 & 0 &  1 & R_{V,2} & \hdots & 0 & 0\\
\vdots  & \vdots & \vdots & \vdots & \hdots & \vdots & \vdots \\
0 & 0 &  1 & R_{V,N} & \hdots & 0 & 0\\
\vdots  & \vdots & \vdots & \vdots & \ddots & \vdots & \vdots \\
0 & 0 & 0 & 0 & \hdots & 1 & R_{V,1} \\
0 & 0 & 0 & 0 & \hdots & 1 & R_{V,2} \\
\vdots  & \vdots & \vdots & \vdots & \hdots & \vdots & \vdots \\
0 & 0 & 0 & 0 & \hdots & 1 & R_{V,N}  \\
1 & R_{I,1} & 0 & 0 & \hdots & 0 & 0\\
1 & R_{I,2} & 0 & 0 & \hdots & 0 & 0\\
\vdots  & \vdots & \vdots & \vdots & \hdots & \vdots & \vdots \\
1 & R_{I, N} & 0 & 0 & \hdots & 0 & 0\\ 
\vdots  & \vdots & \vdots & \vdots & \ddots & \vdots & \vdots \\
0 & 0 & 0 & 0 & \hdots & 1 & R_{I,1} \\
0 & 0 & 0 & 0 & \hdots & 1 & R_{I,2} \\
\vdots  & \vdots & \vdots & \vdots & \hdots & \vdots & \vdots \\
0 & 0 & 0 & 0 & \hdots & 1 & R_{I,N}  \\
1 & R_{G_{\rm RP}, 1} & 0 & 0 & \hdots & 0 & 0\\
1 & R_{G_{\rm RP}, 2} & 0 & 0 & \hdots & 0 & 0\\
\vdots  & \vdots & \vdots & \vdots & \hdots & \vdots & \vdots \\
1 & R_{G_{\rm RP}, N} & 0 & 0 & \hdots & 0 & 0\\ 
\vdots  & \vdots & \vdots & \vdots & \ddots & \vdots & \vdots \\
0 & 0 & 0 & 0 & \hdots & 1 & R_{G_{\rm RP}, 1} \\
0 & 0 & 0 & 0 & \hdots & 1 & R_{G_{\rm RP}, 2} \\
\vdots  & \vdots & \vdots & \vdots & \hdots & \vdots & \vdots \\
0 & 0 & 0 & 0 & \hdots & 1 & R_{G_{\rm RP}, N}  \\
\end{bmatrix}_{(N_{\rm t}\times J\times 5, ~2N_{\rm t})} . 
\label{eq:b6}
\end{align}
The weight matrix $\mathbf{C}$ used in the WLS minimization to obtain the values of $\Delta\mu_{0,i}$ and $\Delta E(B-V)_{0,i}$ as well as the corrected multi-phase PL relations has $1/\sigma^{2}_{\lambda,i,j}$ as elements along its diagonal. It is a square matrix of order $(N_{\rm t}\times J\times 5,  ~N_{\rm t}\times J\times 5)$. It is given in the following matrix form:
\clearpage
\begin{align*}
\mathbf{C} = & \begin{bmatrix}
1/\sigma^{2}_{V,1,1} & 0 & 0 & 0 & \hdots & 0 & \hdots & 0 & \hdots & 0\\ 
0 & 1/\sigma^{2}_{V,2,1} & 0 & 0 & \hdots & 0 & \hdots & 0 & \hdots & 0\\
0 & 0 & 1/\sigma^{2}_{V,3,1} & 0 & \hdots & 0 & \hdots & 0 & \hdots & 0 \\ 
\vdots & \vdots & \vdots & \ddots & \vdots & \vdots & \hdots & \vdots & \vdots & \vdots \\ 
0 & 0 & 0 & \hdots & 1/\sigma^{2}_{V,N_{\rm t},J} & 0 & \hdots & 0 & \hdots & 0 \\
0 & 0 & 0 & 0 & \hdots & 1/\sigma^{2}_{I,1,1} & \hdots & 0 & \hdots & 0 \\
\vdots & \vdots & \vdots & \vdots & \hdots & \vdots & \ddots & \vdots & \vdots & \vdots \\ 
0 & 0 & 0 & 0 & \hdots & 0 & \hdots & 1/\sigma^{2}_{I,N_{\rm t},J} & \hdots & 0 \\
\vdots & \vdots & \vdots & \vdots & \hdots & \vdots & \hdots & \vdots & \ddots & \vdots \\ 
0 & 0 & 0 & 0 & \hdots &  0 & \hdots & 0 & \hdots & 1/\sigma^{2}_{G_{\rm RP},N_{\rm t},J} 
\end{bmatrix}.
\label{eq:b7}
\end{align*}    

%%%%%%%%%%%%%%%%%%%%%%%%%%%%%%%%%%%%%%%%%%%%%%%%%%

% Don't change these lines
\bsp % typesetting comment
\label{lastpage}
\end{document}